\documentclass[aps,pra,reprint,amssymb,superscriptaddress,nofootinbib]{revtex4-2} 
\usepackage{amsmath,amssymb,amsfonts}
\usepackage{graphicx}
\usepackage{dcolumn}
\usepackage{bm}
\usepackage{subfigure}
\usepackage{dsfont}
\usepackage{physics}
\usepackage[colorlinks=true,citecolor=blue,linkcolor=blue,urlcolor=blue]{hyperref}
\usepackage[mathlines]{lineno}

\usepackage{float}
\makeatletter
\let\newfloat\newfloat@ltx
\makeatother
\usepackage{algorithm}
\usepackage{algpseudocode}

\usepackage{braket}
\usepackage[dvipsnames]{xcolor}
\usepackage{tabularray}

\newenvironment{psmallmatrix}
  {\left(\begin{smallmatrix}}
  {\end{smallmatrix}\right)}

\definecolor{tms}{rgb}{1,0,0}

\begin{document}

\title{Quantum simulation of general spin-1/2 Hamiltonians with parity-violating fermionic Gaussian states}

\author{Michael Kaicher}
\let\comma,
\affiliation{PASQAL SAS, 24 rue Emile Baudot - 91120 Palaiseau,  Paris, France}

\author{Joseph Vovrosh}
\let\comma,
\affiliation{PASQAL SAS, 24 rue Emile Baudot - 91120 Palaiseau,  Paris, France}

\author{Alexandre Dauphin}
\let\comma,
\affiliation{PASQAL SAS, 24 rue Emile Baudot - 91120 Palaiseau,  Paris, France}

\author{Simon B. J\"ager}
\let\comma,
\affiliation{Physikalisches Institut, University of Bonn, Nussallee 12, 53115 Bonn, Germany}

\begin{abstract}
We introduce equations of motion for a parity-violating fermionic mean-field theory (PV-FMFT): a numerically efficient fermionic mean-field theory based on parity-violating fermionic Gaussian states (PV-FGS). This work provides explicit equations of motion for studying  the real- and imaginary-time evolution of spin-1/2 Hamiltonians with arbitrary geometries and interactions. We extend previous formulations of parity-preserving fermionic mean-field theory (PP-FMFT) by including fermionic displacement operators in the variational Ansatz. Unlike PP-FMFT, PV-FMFT can be applied to general spin-1/2 Hamiltonians, describe quenches from arbitrary initial spin-1/2 product states, and compute local and non-local observables in a straight-forward manner at the same modest computational cost as PP-FMFT---scaling as $\mathcal O(N^3)$ in the worst case for a system of $N$ spins or fermionic modes. We demonstrate that PV-FMFT can exactly capture the imaginary- and real-time dynamics of non-interacting spin-1/2 Hamiltonians.  We then study the post quench-dynamics of the one- and two-dimensional Ising model in presence of longitudinal and transversal fields with PV-FMFT and compute the single site magnetization and correlation functions, and compare them against results from other state-of-the-art numerical approaches. In two-dimensional spin systems, we show that the employed spin-to-fermion mapping can break rotational symmetry within the PV-FMFT description, and we discuss the resulting consequences for the calculated correlation functions. Our work introduces PV-FMFT as a benchmark for other numerical techniques and quantum simulators, and it outlines both its capabilities and its limitations.

\end{abstract}

\date{\today}

\maketitle

\section{Introduction}

Understanding basic principles such as ground state and thermal properties, as well as the time evolution of interacting quantum many-body systems lies at the heart of quantum simulation. Spurred by the progress made in controlling and manipulating atomic, molecular and optical systems ~\cite{Cornish2024,Gross2017quantum,morgado2021quantum,bruzewicz2019trapped}, as well as other platforms such as those based on superconducting circuits~\cite{kjaergaard2020superconducting}, numerical methods whose computational cost scales favorably with  system size $N$ are required. Existing approaches are often characterized as strongly or weakly correlated, depending on the amount of entanglement they can capture and their computational scalability. While strongly correlated methods are indispensable in regimes where quantum states obey area- or volume-law entanglement~\cite{eisert2010colloquium} (at least for dimensions $>1$), it is not always clear a priori whether a given system truly requires such methods, or whether a weakly correlated approach suffices. Moreover, what constitutes a “successful’’ numerical simulation may vary by context: qualitative reproduction of key mechanisms, as in superconductivity~\cite{bardeen1957microscopic}, may be adequate in some settings, while quantitative agreement, as in quantum chemistry~\cite{helgaker2013molecular}, is essential in others.

Tensor-network approaches~\cite{cirac2021matrix,banuls2023tensor}, including matrix product states (MPS)~\cite{schollwoeck2011density}, tree tensor networks~\cite{tagliacozzo2009simulation}, and higher-dimensional tensor network generalizations~\cite{verstraete2004renormalizationalgorithmsquantummanybody}, are powerful strongly correlated methods. However, their accuracy may break down rapidly, for instance in MPS when the required entanglement entropy grows faster than $\mathcal O(\mathrm{log}(D))$, where $D$ is the maximal affordable bond dimension~\cite{Daley2022}. This occurs particularly during real-time evolution where entanglement typically increases rapidly with time following a volume law~\cite{eisert2010colloquium}. Weakly correlated methods, though limited in the range of correlations they can capture, offer a complementary approach with computational costs that scale polynomially with the system size $N$. The ultimate aim of numerical methods for quantum simulation is to balance computational tractability with an accurate and comprehensive description of both static and dynamically generated quantum correlations.

Possibly one of the simplest examples of an efficient simulation method is spin mean-field theory (SMFT), which belongs to the class of methods that do not capture any correlations or entanglement, but nonetheless has been successful in describing certain phase transitions and order parameters~\cite{weiss1907hypothese,landau1937theory}. SMFT is an example of a variational mean-field method, where the dynamics is restricted to the family of spin product states and the equations of motion (EOMs) follow from a variational principle. Going beyond SMFT to describe quantum correlations, weakly correlated methods based on Gaussian variational states~\cite{Shi2018,hackl2021bosonic} can capture limited forms of correlation and entanglement. Their main representatives are bosonic and fermionic mean-field theory (FMFT), whose variational wave functions are bosonic and fermionic Gaussian states (FGS), respectively. Although far less expressive than tensor-network methods, their approximations and limitations are well understood and can therefore allow for a simple physical interpretation of the efficiently simulated numerical results~\cite{hackl2018aspects,rougerie2021scalinglimitsbosonicground,hackl2021bosonic}. Depending on the system and parameter regime, these methods can yield  qualitative and even quantitative agreement with more strongly correlated methods, but of course can become unreliable whenever the main approximations break down. Importantly, Gaussian mean-field approaches remain valuable for identifying beyond-classical-simulatability regimes  where discrepancies between mean-field predictions and experiment or other classical benchmarks may indicate the emergence of genuinely quantum and strongly correlated behavior~\cite{vovrosh2025simulatingdynamicstwodimensionaltransversefield}. This interplay is becoming increasingly central to the assessment of claims of quantum advantage~\cite{Kim2023,tindall2024efficient,king2025,tindall2025dynamicsdisorderedquantumsystems,mauron2025challengingquantumadvantagefrontier} , underscoring the vital importance of developing and improving Gaussian mean-field approaches.

\begin{figure}
    \centering
    \includegraphics[width=0.9\linewidth]{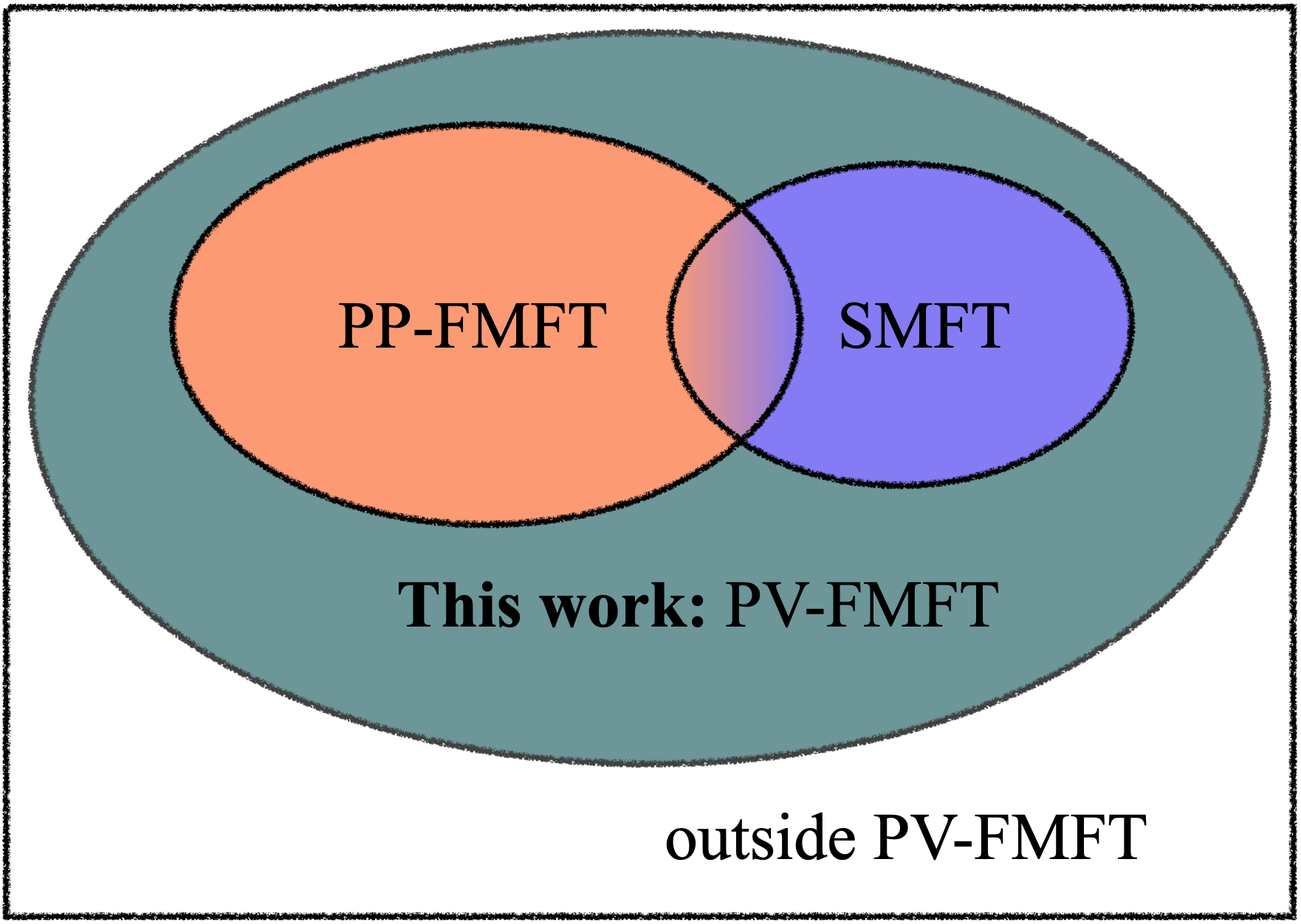}
    \caption{Schematic relation between spin mean-field theory (SMFT) and fermionic mean-field theory (FMFT). Each circle represents a family of Hamiltonians and states which are exactly described by a mean-field theory based on spin-1/2 product states (purple), parity-preserving fermionic Gaussian states (PP-FGS, orange), or parity-violating (PV)-FGS (green), respectively. The vast majority of previous FMFT formulations relied on PP-FGS, restricting applicability to spin systems whose fermionic representation is parity preserving. The PV-FGS we use in this work to formulate a PV-FMFT enable real- and imaginary-time evolution of arbitrary spin-1/2 and fermionic systems at low computational cost.}
    \label{fig:mean_field_separation}
\end{figure}

In this work we derive explicit, numerically stable real- and imaginary-time EOMs for parity-violating fermionic Gaussian states (PV-FGS)---referring to the resulting framework as parity-violating fermionic mean-field theory (PV-FMFT). As an application, we discuss how PV-FMFT applies to general spin-1/2 systems. PV here corresponds to mixing states with even and odd fermionic occupation number. Such mixing appears naturally in bosonic Gaussian states via displacement operators after spin-to-boson mappings such as via a Holstein–Primakoff mapping~\cite{holstein1940field}. In contrast, incorporating PV terms into fermionic Gaussian Ansätze is more subtle, in part due to fermionic superselection rules~\cite{wick1997intrinsic} and in part because PV-FGS involve Grassmann variables~\cite{berazin2012method,barnett2023glaubersudarshanprepresentationsfermions,henderson2024hartree}, for which analytical and numerical tools are less established.

Importantly, although physical fermionic systems conserve parity, spin models mapped to fermions via Jordan–Wigner~\cite{jordan1993paulische} or related encodings~\cite{tranter2015bravyi,BRAVYI2002fermionic} generally yield effective fermionic Hamiltonians containing odd fermionic monomials, and are therefore parity violating. Traditional FMFT approaches restricted to parity-preserving fermionic Gaussian states (PP-FGS) are therefore unable to treat generic spin Hamiltonians under the above mappings. Alternative spin-to-fermion mappings that guarantee PP do exist~\cite{rao2025dynamical,knolle2018dynamics,yilmaz2022phasediagramskitaevmodels}, but require doubling the number of fermionic modes and enforcing gauge constraints throughout the evolution~\cite{KITAEV20062}. In contrast, our approach maps $N$ spins to only $N{+}1$ fermionic modes without additional constraints~\cite{rao2025dynamical}, making PV-FMFT a simple and efficient description for spin and fermionic systems.

While our work provides, to the best of our knowledge, the first explicit real- and imaginary-time EOMs for a general PV-FGS variational Ansatz, it builds on earlier studies of PV fermionic frameworks~\cite{fukutome1977new,fukutome1977properties,moussa2012generalizedunitarybogoliubovtransformation,nishiyama2019remarks,henderson2024hartree,Mirjafarlou2024,lyu2024displacedfermionicgaussianstates}. Following the Gaussian variational formalism of Ref.~\cite{Shi2018}, we derive closed-form EOMs and present stable numerical integration schemes. As summarized in Figure~\ref{fig:mean_field_separation}, PV-FMFT extents and includes parity-preserving fermionic mean-field theory (PP-FMFT), exactly captures ground states and time evolution of non-interacting spin Hamiltonians, and enables efficient  simulations of large spin-1/2 and arbitrary fermionic systems.

The paper is organized as follows. Section~\ref{theory} introduces the theoretical PV-FMFT framework, in particular,  Section~\ref{pv_fmft} presents the explicit EOMs. The numerical results are shown in Section~\ref{num_results}, including simulations of non-interacting spins in Section~\ref{non_interacting_numerics} and the one- and two-dimensional Ising model in presence of longitudinal and transversal fields in Section~\ref{tfim}. Section~\ref{summary} concludes with a summary and an outlook. The main proofs and calculations of this work are presented in the Appendix.

\section{Theory\label{theory}}

\subsection{Parity-preserving fermionic mean-field theory\label{pp-fmf}}
 We define the fermionic creation and annihilation operators $\hat c_p^\dag$ and $\hat c_p$, which satisfy the canonical anti-commutation relations $\{\hat c_p,\hat c_q^\dag\}=\hat c_p\hat c_q^\dag + \hat c_q^\dag\hat c_p=\delta_{pq}$, where $\delta_{pq}=1$ if $p=q$ and $\delta_{pq}=0$ if $p\neq q$. In this paper, we describe fermionic systems by introducing Majorana operators $\hat a_{2p} = \hat c_p^\dag + \hat c_p$ and $\hat a_{2p+1} = i(\hat c_p^\dag -\hat c_p)$, which by definition are Hermitian operators satisfying $\{\hat a_l,\hat a_m\}=2\delta_{lm}$. We also introduce Majorana operators with a permuted numbering which are in this paper denoted with capital letters, $\hat A_0=\hat a_0,\hat A_1=\hat a_2, \dots, \hat A_{N-1}=\hat a_{2N-2}$ and $\hat A_{N}=\hat a_1,\hat A_{N+1}=\hat a_3, \dots, \hat A_{2N-1}=\hat a_{2N-1}$, for a system of $N$ fermionic modes. The introduction of both notations is purely technical. The ordering $\hat{\mathbf a}=(\hat a_0,\ldots,\hat a_{2N-1})^{T}$ is convenient for the Jordan–Wigner transformation discussed later, whereas the ordering $\hat{\mathbf A}=(\hat A_0,\ldots,\hat A_{2N-1})^{T}$ is better suited for certain algebraic manipulations and follows the notation of Ref.~\cite{Shi2018}.
 
 Any fermionic Hamiltonian that can be written as an even polynomial of fermionic operators, 
\begin{align}
    \hat H = \text{poly}_{\text{even}}(\hat {\bf a}), \label{ppham}
\end{align}
satisfies the superselection rule and commutes with the parity operator
\begin{align}
    \hat \Pi = e^{i\pi\sum_{p=0}^{N-1}\hat n_p} = (-i)^{N}\hat a_0\hat a_1\cdots\hat a_{2N-2},\label{parity_op}
\end{align}
where $\hat n_p=\hat c_p^\dag \hat c_p$ is the occupation-number operator. The eigenvalues of the parity operator in Eq.~\eqref{parity_op} are given by $+1$ and $-1$ and quantum states which are eigenstates of the parity operator with eigenvalue $+1$ ($-1$) are said to have even (odd) parity. 

Since physical fermionic Hamiltonians are of the form of Eq.~\eqref{ppham}, it is not surprising that wave-function-based many-body methods aimed at describing fermionic systems have focused on state Ans\"atze which also commute with the parity operator. A prominent PP fermionic wave function  is given by the PP-FGS, defined as
\begin{align}
    \ket{\Psi_{\text{FGS}}} = \hat U_{\text{FGS}}\ket{\mathbf 0},\label{fgs2}
\end{align}
where $\ket{\mathbf 0} = \ket{0}^{\otimes N}$ denotes the fermionic vacuum state and  we defined the unitary operator
\begin{align}
    \hat U_{\text{FGS}} = e^{\frac{i}{4}\mathbf{\hat A}^T\boldsymbol\xi\mathbf{\hat A}},\label{fgs1}
\end{align}
where $\boldsymbol{\xi}$ is a $(2N\times 2N)$ skew-symmetric and Hermitian matrix. Here, we introduced the short-hand notation $\sum_{jk}\xi_{jk}\hat A_j\hat A_k=\mathbf{\hat A}^T\boldsymbol{\xi}\mathbf{\hat A}$. From the definition of Eq.~\eqref{fgs1} it can be seen that Ansatz in Eq.~\eqref{fgs2} describes a PP wave function since it is generated by an even polynomial of fermionic operators and acts on an initial state with fixed (in this case even) parity, which cannot mix the parity sectors. The unitary in Eq.~\eqref{fgs1} transforms  Majorana operators as follows,
\begin{align}
    \hat U_{\text{FGS}}^\dag \mathbf{\hat A} \hat U_{\text{FGS}}=&\mathbf R\mathbf{\hat A},\label{traf1}
\end{align}
where $\mathbf R =e^{i\boldsymbol{\xi}}$ describes an orthogonal rotation. The central object of FMFT is the real-valued and skew symmetric covariance matrix $\boldsymbol{\Gamma}$, whose matrix elements are given by 
\begin{align}
\Gamma_{pq} =& \frac{i}{2}\braket{[\hat A_p,\hat A_q]}_{\text{FGS}},\label{fgs14}
\end{align}
where $[\hat A_p,\hat A_q]=\hat A_p\hat A_q-\hat A_q\hat A_p$, and we introduced a short-hand notation for expectation values $\braket{[\hat A_p,\hat A_q]}_{\text{FGS}}=\bra{\Psi_{\text{FGS}}}[\hat A_p,\hat A_q]\ket{\Psi_{\text{FGS}}}$. 
The variational wave function parameters $\boldsymbol{\xi}$ are related to the covariance matrix through 
\begin{align}
    \boldsymbol{\Gamma} =& -\mathbf R \boldsymbol{\Upsilon} \mathbf R^T,
\end{align}
where 
\begin{align}
\boldsymbol{\Upsilon} =& \begin{pmatrix}
        \mathbf 0_N & \mathbf 1_N\\
        -\mathbf 1_N & \mathbf 0_N
    \end{pmatrix},\label{fgs13}
\end{align}
and $\mathbf 1_N$ and $\mathbf 0_N$ denotes the $(N\times N)$ identity and zero matrices, respectively. FGS satisfy Wick's theorem, which states that expectation values of even monomials of fermionic operators can be computed efficiently by the Pfaffian of submatrices of the covariance matrix \cite{bach1994generalized}
\begin{align}
    \braket{\hat A_{i_1}\hat A_{i_2}\cdots \hat A_{i_m}}_{\text{FGS}}=(-i)^{\frac{m}{2}}\text{Pf}\left(\left.\boldsymbol{\Gamma}\right|_{i_1,i_2,\dots,i_m}\right),\label{wick}
\end{align}
where the shorthand notation $\left.\boldsymbol{\Gamma}\right|_{i_1,i_2,\dots,i_m}$ denotes the submatrix of $\boldsymbol{\Gamma}$ only containing the rows and columns described by the ordered  index set $\{i_1,i_2,\dots,i_m\}$, and $\text{Pf}(\mathbf X)$ denotes the Pfaffian of a real and skew symmetric matrix $\mathbf X$ \cite{Wimmer2012}. The Pfaffian of an $(m\times m)$-matrix is always zero if $m$ is an odd integer, which means that Eq.~\eqref{wick} is only nonzero for even-degree monomials $m=2n$, with $n\in\mathds N, n\leq N$. The computational cost for computing the Pfaffian of a $(m\times m)$-matrix scales as $\mathcal O(m^3)$, which makes the evaluation of expectation values---such as the expectation value of a local Hamiltonian of the form of Eq.~\eqref{ppham}---efficient. 

In the following we describe the well-studied mean-field method 
PP-FMFT. As detailed in Ref.~\cite{Shi2018}, the variational Ansatz in Eq.~\eqref{fgs2} can be used to approximate the ground-state\footnote{It is also possible to describe finite temperature properties of fermionic systems within the Gaussian approximation, see e.g. Refs.~\cite{shi2020finite, kraus2010generalized}.} or real time evolution of a fermionic or spin system. All information about the FGS is contained in   $\boldsymbol{\Gamma}$ (or alternatively in $\boldsymbol{\xi}$). In the derivation of the EOMs for $\boldsymbol{\Gamma}$, one makes use of the the following normal-ordering expansion: For any operator $\hat \Xi$ that is an even polynomial of Majorana operators, its normal-ordering expansion up to quadratic order is given by
\begin{align}
    \hat U_{\text{FGS}}^\dag\hat \Xi\hat U_{\text{FGS}} = \langle\hat \Xi\rangle_{\text{FGS}} + \frac{i}{4}:\hat{\mathbf A}^T \mathbf R^T\boldsymbol \Xi_m\mathbf R \hat{\mathbf A}: + \delta\hat \Xi.\label{a12}
\end{align}
The above expansion follows from Wick's theorem, and $\delta\hat \Xi$ denotes a normal-ordered even polynomial of Majorana operators of order four or higher,  $:\hat O:$ denotes normal-ordering of a fermionic operator $\hat O$, and we defined the matrix 
\begin{align}
\left(\Xi_m\right)_{pq}=4\frac{d\langle\hat \Xi\rangle_{\text{FGS}}}{d\Gamma_{pq}}.\label{a13}
\end{align} 

The imaginary time EOM of the covariance matrix is given by 
\begin{align}
    \frac{d\boldsymbol{\Gamma}}{d\tau} 
    =&-\mathbf H_m -\boldsymbol{\Gamma}\mathbf H_m\boldsymbol{\Gamma},\label{b8}
\end{align}
where $\tau$ denotes the imaginary time and 
\begin{align}
    \mathbf H_m=4\frac{d\braket{\hat H}_{\text{FGS}}}{d\boldsymbol{\Gamma}}.
\end{align}
The solution to Eq.~\eqref{b8} describes the ground state within the Gaussian approximation [see Eq.~\eqref{a12}] of any PP Hamiltonian $\hat H$ as defined in Eq.~\eqref{ppham} within the family of PP-FGS defined in Eq.~\eqref{fgs2}.

A similar EOM can be derived which describes the real-time evolution of a PP Hamiltonian as defined in Eq.~\eqref{ppham} within the Gaussian approximation,
\begin{align}
    \frac{d\boldsymbol{\Gamma}}{dt}=&\left[\mathbf H_m, \boldsymbol{\Gamma}\right].\label{c8}
\end{align}
The solution of Eq.~\eqref{c8} describes the real-time evolution and can be used to describe e.g. the post-quench dynamics of a PP Hamiltonian under the Gaussian approximation from an initial state, provided that the latter can be represented as a PP-FGS. 

\subsection{Parity-violating fermionic mean-field theory\label{pv_fmft}}

\subsubsection{Motivation}
While many realistic fermionic Hamiltonians can be represented as even polynomials as in Eq.~\eqref{ppham} this is not true for every Hamiltonian. An important class are generic spin-1/2 Hamiltonians. One can map any spin-1/2 Hamiltonian onto a fermionic Hamiltonian by an exact mapping such as the Jordan-Wigner (JW) transformation ~\cite{jordan1993paulische}. However, when mapping the spin Hamiltonian to a fermionic Hamiltonian via JW, this generally results in a polynomial with even and odd monomials
\begin{align}
    \hat H = \text{poly}(\hat {\bf a}).\label{pvham}
\end{align}
This means that the fermionic parity can be broken (due to the presence of odd monomials of fermionic operators). As a consequence, the FMFT applied to spin Hamiltonians requires the consideration of the larger family of PV-FGS, as illuatrated in Figure~\ref{fig:mean_field_separation}. 

In what follows, we will show how including the PV fermionic displacement operator to the variational Ansatz of Eq.~\eqref{fgs2}, followed by the introduction of an auxiliary mode-mediated mapping, leads to a consistent set of EOMs of the same form as Eqs.~\eqref{b8}-\eqref{c8}. This allows us to derive a FMFT that can be applied to any spin-1/2- and PV fermionic Hamiltonian. In the context of simulating quantum quenches of arbitrary spin-1/2 Hamiltonians, our approach further allows to include quantum quenches of arbitrary spin-1/2 Hamiltonians, which for instance allows us to start from arbitrary uncorrelated spin states.

\subsubsection{The appearance of parity-violating fermionic Hamiltonians}
One can map $N$ spinless fermions to $N$ spin-1/2 systems and vice versa using a number of different mappings \cite{jordan1993paulische,BRAVYI2002fermionic,seeley2012bravyi,Jiang2020optimalfermionto,tranter2015bravyi}. In this work we consider the JW transformation~\cite{jordan1993paulische}, which is given by
\begin{align}
    \hat \sigma^x_p \rightarrow&\hat S_{0,p-1} \hat a_{2p}\label{j2},\\
    \hat \sigma^y_p \rightarrow& \hat S_{0,p-1}\hat a_{2p+1}\label{j3},\\
    \hat \sigma^z_p \rightarrow& \mathds 1 -2\hat c_p^\dag \hat c_p=\hat S_{p,p},\label{j4}
\end{align}
where we defined the fermionic string operator 
\begin{align}
    \hat S_{p,q} =\left(\prod_{l=p}^{q}(-i)\hat a_{2l}\hat a_{2l+1}\right), \label{string}
\end{align}
and where $\hat\sigma_k^\alpha$ with $\alpha\in\{x,y,z\}$ denotes a Pauli operator. For all systems in this manuscript, we choose a snake ordering\footnote{The choice of the labeling has a measurable effect on the FMFT performance \cite{henderson2024fermionic,henderson2023restoringpermutationalinvariancejordanwigner}.} of the  lattices for the FMFT. 

We will study two examples of spin-1/2 systems which result in PV fermionic Hamiltonians of the form of Eq.~\eqref{pvham} under the JW mapping. To demonstrate the validity of our approach, we first study the simple example of a non-interacting spin-1/2 Hamiltonian 
\begin{align}
    H = \sum_p\frac{J_p^x}{2}\hat\sigma_p^x + \sum_p\frac{J_p^y}{2}\hat\sigma_p^y+\sum_p\frac{J_p^z}{2}\hat\sigma_p^z,\label{nis1}
\end{align}
where $J_p^{\alpha}$ are random real-valued coefficients. After the JW transformation, the Hamiltonian of Eq.~\eqref{nis1} is given by
\begin{align}
    H =& \sum_{p=1}^N\left(\frac{J_p^x}{2}\hat S_{1,p-1}\hat a_{2p} +\frac{J_p^y}{2}\hat S_{1,p-1}\hat a_{2p+1} +\frac{J_p^z}{2}\hat S_{p,p}\right).\label{nis2}
\end{align}
Note, that we have chosen to use indices which start at 1 instead of 0 in  the fermionic representation since the 0-mode will be used for an auxiliary mode introduced later on. 

While the Hamiltonian of Eq.~\eqref{nis1} can be solved exactly by SMFT, as mentioned for instance in Ref.~\cite{ryabinkin2018relation}, the PP-FMFT introduced in Section~\ref{pp-fmf} cannot solve this simple spin system. This can be traced back  to the presence of PV terms (i.e. odd fermionic monomials) in its fermionic representation in Eq.~\eqref{nis2} as well as the presence of monomials of higher than quadratic order. While we will introduce PV-FGS in the next subsection, it should be mentioned here that it was shown in Refs.~\cite{henderson2024hartree,lyu2024displacedfermionicgaussianstates} that any spin product state (which forms the variational Ansatz of SMFT) can be expressed as a PV-FGS. Therefore, a PV-FMFT should be able to reproduce the exact ground state and dynamics of Eq.~\eqref{nis1}, which makes the non-interacting spin Hamiltonian a perfect proof-of-principle system for our purpose. 

As a non-trivial spin-1/2 example, we consider the Transverse Field Ising Model (TFIM), 
\begin{align}
    H 
    =&\frac{1}{8}\sum_{k\neq l}J_{kl}\hat \sigma^z_k\hat \sigma^z_l + \frac{1}{2}\sum_k \Omega_k\hat \sigma^x_k+ \frac{1}{2}\sum_k \zeta_k\hat \sigma^z_k,\label{pb1}
\end{align}
where $J_{kl}=J_{lk}\in\mathds R$ (with $J_{kk}=0$) describes the interaction, $\Omega_k$ and $\zeta_k$ describe transversal and longitudinal magnetic fields, respectively.  By `non-trivial', we here mean that we do not expect FMFT to describe the ground-state or time evolution of the respective Hamiltonian exactly, with the exception of a few special cases. Under the JW transformation, the Hamiltonian of Eq.~\eqref{pb1} results in
\begin{align}
    H =&-\frac{1}{8}\sum_{k\neq l}J_{kl}\hat a_{2k}\hat a_{2k+1}\hat a_{2l}\hat a_{2l+1} \nonumber\\
    &+ \frac{1}{2}\left(\sum_{k=1}^N\Omega_k \hat S_{1,k-1}\hat a_{2k}+\zeta_k\hat S_{k,k}\right),
    \label{pb1b}
\end{align}
which is PV due to the presence of the odd fermionic monomials.

In the absence of a longitudinal field component, $\zeta_k=0$, the Hamiltonian in Eq.~\eqref{pb1} can be rotated into a basis where the interaction is described by the operator $\hat \sigma^x_k\hat \sigma^x_l$ (which we refer to as the `$XX$-representation'), in which case the resulting fermionic Hamiltonian under the JW mapping is of the form of Eq.~\eqref{ppham}. In this special case, one can use the formalism of Section~\ref{pp-fmf} to describe ground state properties, or even study quantum quenches from the field-polarized state (in fact any initial state  which can be represented by a PP-FGS). In the case of a one-dimensional spin chain and only nearest-neighbor interactions,  as well as a vanishing longitudinal field, the ground state of Eq.~\eqref{pb1} is solved exactly by PP-FMFT. Even for interactions in the strong long-range regime (for $\alpha\geq 1$ which describes the scaling of the interaction with Euclidean distance \(J_{kl}=r_{kl}^{-\alpha}\)), FGS can provide a good approximation of the ground state energy and entanglement in absence of a longitudinal field, as shown in Ref.~\cite{Kaicher2023}. It is known that the presence of the longitudinal field breaks integrability of the one-dimensional TFIM~\cite{kim2014testing,Kormos2017}. In the $XX$-representation, the TFIM Hamiltonian is given by 
\begin{align}
    H 
    =&\frac{1}{8}\sum_{k\neq l}J_{kl}\hat \sigma^x_k\hat \sigma^x_l + \frac{1}{2}\sum_k \Omega_k\hat \sigma^z_k+ \frac{1}{2}\sum_k \zeta_k\hat \sigma^x_k, \label{hd1}
\end{align}
where the last term describing the longitudinal field $\boldsymbol{\zeta}$ is responsible for the breaking of the fermionic parity. 

When one is interested in applying the PP-FMFT to general spin 1/2-Hamiltonians [or in case of the TFIM of Eq.~\eqref{pb1}, an instance where the longitudinal and transversal field are non-zero] and potentially consider initial states which are not described by a PP-FGS, the PP-FMFT described in Section~\ref{pp-fmf} is no longer suitable. The latter case for instance arises when one wants to study quenches from a transverse-field polarized state, and a state that is polarized along the direction of the interaction. Numerically, one can still apply PP-FMFT, but one will not be able to describe all Hamiltonian terms since expectation values of odd monomials of fermionic operators w.r.t. a PP-FGS vanish, see Eq.~\eqref{wick}, and because only a small group of spin product states can be expressed as PP-FGS. Therefore, in order to treat general spin-1/2 systems with FMFT, one has to extend the variational FGS Ansatz of Eq.~\eqref{fgs2} in a way that naturally allows to break the fermionic parity.  This will be the content of the next subsection. 

\subsubsection{Parity-violating fermionic Gaussian states}
We define the complex-valued vectors $\mathbf u=(u_1,\dots,u_N)^T$ and $\mathbf v=(v_1,\dots,v_N)^T$, and the $(2N\times 2N)$-matrix $\mathbf M$. We assume that $\mathbf{JM}$ is a complex and skew-symmetric matrix, $\mathbf{JM}=-(\mathbf{JM})^T=-\mathbf M^T\mathbf J^T$, where we defined
\begin{align}
    \mathbf J=\begin{pmatrix}
        \mathbf 0_N&\mathbf 1_N\\
        \mathbf 1_N&\mathbf 0_N
    \end{pmatrix},\label{npv_fgs_7}
\end{align}
and define $\ket{\mathbf I} = \ket{i_1,i_2,\dots,i_N}$, with $i_k\in\{0,1\}$. We will also define a vector of fermionic annihilation operators $\hat{\mathbf c} = (\hat c_1, \hat c_2,\dots,\hat c_N)^T$ and use $(\hat{\mathbf c}^\dag \ \hat{\mathbf c})$ in slight abuse of notation to denote a row vector of all creation and annihilation operators. 

In the spirit of the variational Ansatz of Eq.\eqref{fgs2}, we will formulate a PV-FMFT based on PV-FGS. More precisely, we consider the operator  
\begin{align}
    \hat{\mathcal F}(\mathbf M,\mathbf u,\mathbf v)=& \exp\left( \frac{1}{2} \begin{pmatrix} \hat{\mathbf c}^\dag & \hat{\mathbf c}\end{pmatrix} \mathbf M \begin{pmatrix}
        \hat{\mathbf c} \\ \hat{\mathbf c}^\dag
    \end{pmatrix} +\mathbf u^\dag \hat{\mathbf c}^\dag +\mathbf v^T\hat{\mathbf c} \right),\label{npv_fgs_36}
\end{align}
which allows us to define a PV-FGS~\cite{Mirjafarlou2024}
\begin{align}
\ket{\tilde \Psi_{\text{PV}}(\mathbf{M},\mathbf{u},\mathbf{v};\mathbf{I})}
    =& \hat{\mathcal F}(\mathbf M,\mathbf u,\mathbf v)\ket{\mathbf I},\label{npv_fgs_37}
\end{align}
where the linear operators (which depend on $\mathbf u$ and $\mathbf v$) are responsible for the breaking of the fermionic parity. The `tilde' indicates that Eq.~\eqref{npv_fgs_36} is not necessarily a unitary operator, and therefore $\ket{\tilde \Psi_{\text{PV}}}$ not necessarily a normalized state vector. 

While in principle one can formulate a PV-FMFT based on Eq.~\eqref{npv_fgs_37}, the fact that we are now dealing with a non-linear transformation in the Hilbert space of $N$ fermionic modes would make a derivation quite complicated. Instead, we will show in the following that a simple mapping to a Hilbert space containing $N+1$ fermionic modes will lead to a linear canonical transformation. Thus, the PV-FMFT in $\mathcal H^{(N)}$ can be formulated as a PP-FMFT in $\mathcal H^{(N+1)}$ as introduced in Section~\ref{pp-fmf}. 

\subsubsection{The Colpa mapping}
\begin{figure*}
    \centering
    \includegraphics[width=0.9\linewidth]{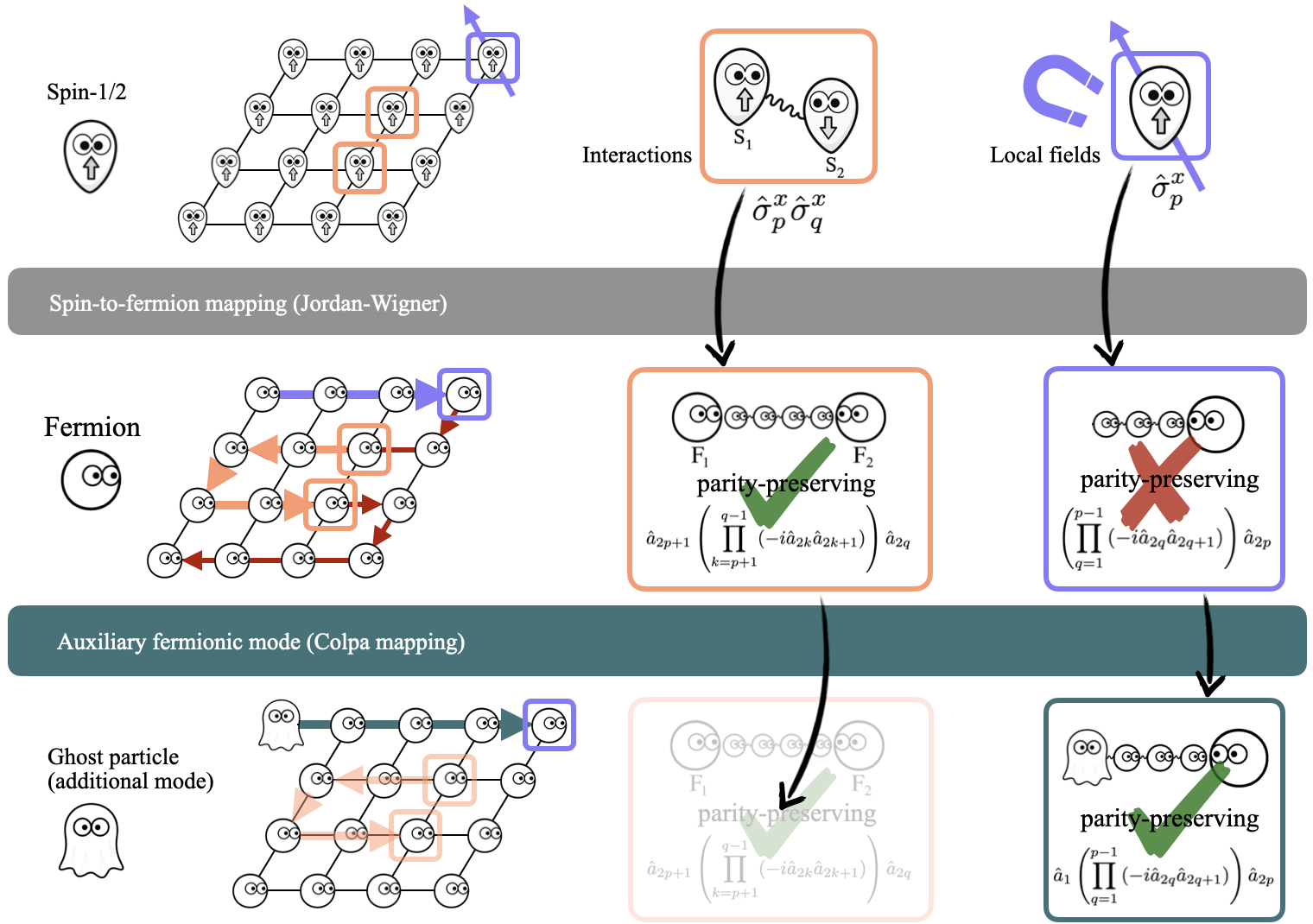}
    \caption{Workflow diagram describing the steps to represent an arbitrary spin-1/2 Hamiltonian (here a two-dimensional square lattice) as a PP fermionic Hamiltonian. We consider the TFIM model where the spin-1/2 Hamiltonian is mapped to its fermionic description via the Jordan-Wigner transformation. While interaction terms (orange box) result in PP fermionic operators, local fields can lead to PV terms (blue box), which cannot be described by PP-FMFT. By introducing a single auxiliary mode (`ghost particle') we can use the Colpa mapping: previously odd fermionic monomials are turned to even fermionic monomials (dark green box), while previously even fermionic monomials are left invariant. The  PV Hamiltonian prior to the mapping is therefore turned into a PP Hamiltonian, for which we formulate a FMFT based on a linear combination of two FGS of opposite parities. Thick colored arrows in the bottom two lattices represent the fermionic modes which are involved when describing the single-spin as well as spin-interaction terms under the Jordan-Wigner and Colpa mapping, assuming snake ordering (red small arrows) of the lattice.}
    \label{fig:workflow}
\end{figure*}

We can map any PV-Hamiltonian as in Eq.~\eqref{pvham} to a PP-Hamiltonian of the form of Eq.~\eqref{ppham}, and any PV-FGS as in Eq.~\eqref{npv_fgs_37} to a linear combination of two PP-FGS by introducing an auxiliary mode (originally referred to as a `ghost particle'). We will reserve the mode $0$---i.e. Majorana operators $\hat a_0$ and $\hat a_1$---for this ghost particle and define the Colpa mapping~\cite{colpa1979diagonalisation},
\begin{align}
    \hat a_{2j}&\stackrel{\text{Colpa}}{\rightarrow}-i\hat a_1\hat a_{2j}\label{colpa3}\\
    \hat a_{2j+1}&\stackrel{\text{Colpa}}{\rightarrow} -i\hat a_1\hat a_{2j+1},\label{colpa4}
\end{align}
where $j\in\{1,\dots,N\}$ denotes the modes of the original Hamiltonian. The absence of the operator $\hat a_0$ in Eqs.~\eqref{colpa3}-\eqref{colpa4} will become important when deriving the EOMs of the PV-FMFT later on. The Colpa mapping has the property that even monomials of fermionic operators remain unaffected, while odd monomials are turned into even monomials, i.e. 
\begin{align}
    \hat H^{(N)} = \text{poly}(\hat a) \ \stackrel{\text{Colpa}}{\rightarrow} \ \hat H'{}^{(N+1)} = \text{poly}_{\text{even}}(\hat a),
\end{align}
where the superscript indicates that the left (right)-hand side Hamiltonian is described by $N$ ($N+1$) fermionic modes, and the prime is used in order to distinguish operators and states before and after the mapping. Figure~\ref{fig:workflow} summarizes the main steps leading from the initial spin-1/2 Hamiltonian to a PP fermionic Hamiltonian through the Colpa mapping. 

The extended Hilbert space with $N+1$ modes $\mathcal H^{(N+1)}$ can be projected on to the subspace of the original Hilbert space $\mathcal H^{(N)}$ using the following mapping~\cite{colpa1979diagonalisation}
\begin{align}
    \ket{\mathbf I}\stackrel{\text{Colpa}}{\rightarrow} \ket{C(\mathbf{I})'}=\frac{1}{\sqrt{2}}\left(\ket{0,\mathbf I} + \ket{1,\mathbf I}\right),\label{npv_fgs_53}
\end{align}
which projects onto the $\hat \sigma^x$ eigenstate of the ghost particle mode $0$. Thus, the PV-FGS defined in Eq.~\eqref{npv_fgs_37} is mapped to a linear combination of two PP-FGS, 
\begin{align}
\ket{\tilde \Psi_{\text{PV}}(\mathbf{M},\mathbf{u},\mathbf{v};\mathbf{I})}  \stackrel{\text{Colpa}}{\rightarrow} &   \mathcal F(\mathbf{M'})\ket{C(\mathbf{I})'}\label{pvngs}
\end{align}
where we defined the even fermionic Gaussian operator
\begin{align}
    \mathcal F(\mathbf{M'})=\exp\left( \frac{1}{2} \begin{pmatrix} \hat{c}_0^\dag&\hat{\mathbf{ c}}^\dag & \hat{c}_0& \hat{\mathbf{ c}} \end{pmatrix} \mathbf M' \begin{pmatrix}
        \hat{c}_0\\ \hat{\mathbf{ c}} \\ \hat{c}_0^\dag\\\hat{\mathbf{ c}}^\dag
    \end{pmatrix} \right)\label{npv_fgs_50}
\end{align}
and, writing the original matrix of Eq.~\eqref{npv_fgs_36} as $\mathbf M = \left(\begin{smallmatrix}\mathbf M^{(11)}&\mathbf M^{(12)}\\\mathbf M^{(21)}&\mathbf M^{(22)}\end{smallmatrix}\right)$, we define 
\begin{align}
    \mathbf{M'} = \begin{pmatrix}
        0 & \mathbf v^T & \mathbf 0 & \mathbf u^\dag \\
        \mathbf u^* & \mathbf M^{(11)} & -\mathbf u^* & \mathbf M^{(12)}\\
        0 & -\mathbf v^T & \mathbf 0 & -\mathbf u^\dag \\
        \mathbf v & \mathbf M^{(21)} & -\mathbf v & \mathbf M^{(22)}
    \end{pmatrix}.\label{npv_fgs_51}
\end{align}
Note, that the fermionic creation and annihilation operators in Eq.~\eqref{npv_fgs_50} now include the auxiliary mode $0$. 

The fermionic Hamiltonian of Eqs.~\eqref{nis2} which describes the random non-interacting spin Hamiltonian of Eqs.~\eqref{nis1} results under the Colpa mapping in 
\begin{align}
    H' =& \frac{1}{2}\sum_{p=1}^N\left[-i\hat a_1\hat S_{1,p-1}(J_p^x\hat a_{2p}+J_p^y\hat a_{2p+1})+\hat J_p^zS_{p,p}\right].\label{nis3}
\end{align}
The fermionic Hamiltonian of Eqs.~\eqref{pb1b} which describes the TFIM  Hamiltonian of Eqs.~\eqref{pb1} results under the Colpa mapping in 
\begin{align}
    H'=&-\frac{1}{8}\sum_{\substack{k,l=1 \\ k\neq l}}J_{kl}\hat a_{2k}\hat a_{2k+1}\hat a_{2l}\hat a_{2l+1} \nonumber\\&- \frac{i}{2}\sum_{k=1}^N\Omega_k\hat a_1 \hat S_{1,k-1}\hat a_{2k}+\frac{1}{2}\sum_{k=1}^N\zeta_k\hat S_{k,k}.\label{pb4}
\end{align}
Thus, we see that both spin-Hamiltonians which in their original JW transformed formulation where of the PV form of Eq.~\eqref{pvham} are now of the PP form of Eq.~\eqref{ppham}.

\subsubsection{Equations of motion of the parity-violating fermionic mean-field theory}
The goal of this work is to present EOMs which allow to describe PV fermionic Hamiltonians with a FMFT based on FGS. In this subsection, we will present our main result: By restricting Eq.~\eqref{npv_fgs_36} to unitary transformations, using the strategy developed in Ref.~\cite{Shi2018}, we can show that the EOMs for a PV-FMFT are analogue to Eqs.~\eqref{b8}-\eqref{c8}. 

We consider the following unitary PV-FGS Ansatz
\begin{align}
    \ket{\Psi_{\text{PV}}'} =& \hat U_{\text{FGS}}'\ket{C(\mathbf 0)'},\label{ppprime}
\end{align}
where $\ket{C(\mathbf 0)'}=\frac{1}{\sqrt{2}}\left(\ket{0,0^{\otimes N}} + \ket{1,0^{\otimes N}}\right)$ follows from Eq.~\eqref{npv_fgs_53}. Strictly speaking, Eq.~\eqref{ppprime} is a sum of two FGS and thus no longer a FGS, but we will still call it a PV-FGS in slight abuse of language. Just like in Eq.~\eqref{fgs1}, we have a PP-FGS generator in the Hilbert space $\mathcal H^{(N+1)}$
\begin{align}
    \hat U_{\text{FGS}}' = e^{\frac{i}{4}\mathbf A'^T\boldsymbol{\xi}'\mathbf A'},\label{npf_fgs_89}
\end{align}
where 
\begin{align}
    \boldsymbol{\xi'} =-\frac{i}{2}(\mathbf W')^*\mathbf J'\mathbf M'(\mathbf W')^\dag\label{npf_fgs_90}
\end{align}
describes a purely imaginary and skew-symmetric matrix, and $\mathbf M'$ in Eq.~\eqref{npf_fgs_90} is a special case of Eq.~\eqref{npv_fgs_51}, 
\begin{align}
    \mathbf{M'} = \begin{pmatrix}
        0 & -\mathbf u^T & 0 & \mathbf u^T \\
        \mathbf u & \mathbf M^{(11)} & -\mathbf u & \mathbf M^{(12)}\\
        0 & \mathbf u^T & 0 & -\mathbf u^T \\
        -\mathbf u & \mathbf M^{(21)} & \mathbf u & \mathbf M^{(22)}
    \end{pmatrix},\label{npv_fgs_88}
\end{align}
which follows from the unitarity condition $\mathbf v = -\mathbf u$. We further introduced the matrix which describes how to transform between fermionic creation and annihilation operators and Majorana operators, \begin{align*}
    \hat{\mathbf A}'=\mathbf W'\begin{pmatrix}
    \hat{c}_0\\
        \hat{\mathbf c} \\ \hat{c}_0^\dag\\
        \hat{\mathbf c}^\dag
\end{pmatrix},
\end{align*} where 
\begin{align}
    \mathbf W' = \begin{pmatrix}
        \mathds 1_{N+1}&\mathds 1_{N+1} \\
        -i\mathds 1_{N+1} &i\mathds 1_{N+1} 
    \end{pmatrix}.\label{fgs16}
\end{align}
Similar to Eq.~\eqref{traf1}, Eq.~\eqref{npf_fgs_89} describes a linear canonical transformation of the fermionic operators in the extended Hilbert space $\mathcal H^{(N+1)}$, $ \hat U_{\text{FGS}}'{}^\dag \mathbf{\hat A}' \hat U_{\text{FGS}}'=\mathbf R'\mathbf{\hat A}'$, where $\mathbf R'=e^{i\boldsymbol \xi'}$ and $\text{det}(\mathbf R')=1$. Thus, using the Colpa mapping has simplified the transformation of the PV-generator from a quadratic canonical transformation in $\mathcal H^{(N)}$ to a PP-generator describing a linear canonical transformation in the extended Hilbert space $\mathcal H^{(N+1)}$. Analog to Eq.~\eqref{fgs14}, we define the covariance matrix $\boldsymbol{\Gamma'}$ and mean-field matrix $\mathbf{H}_m'=4\frac{d\braket{\hat H'}_{\text{PV}}}{d\boldsymbol{\Gamma'}}$ in the extended Hilbert space $\mathcal H^{(N+1)}$. 

The EOMs that describe the imaginary time evolution of  the PV-FGS of Eq.~\eqref{ppprime}  is given by 
\begin{align}
	\frac{d\boldsymbol{\Gamma}'}{d\tau} 
	=&-\mathbf{H}_m' -\boldsymbol{\Gamma}'\mathbf{H}_m' \boldsymbol{\Gamma}'.\label{eom_ite}
\end{align}
while the real-time evolution is given by
\begin{align}    
	\frac{d\boldsymbol{\Gamma}'}{dt} 
	=& [\mathbf{H}_m', \boldsymbol{ \Gamma}'].\label{eom_rte}
\end{align}
A detailed proof for the validity of Eqs.~\eqref{eom_ite}-\eqref{eom_rte} is provided in Appendix~\ref{proof_eom}. The computational cost for simulating PV-FMFT is asymptotically the same as simulating a PP-FMFT, since only a single auxiliary fermionic mode has to be added, with the most costly numerical operation being the evaluation of the Pfaffian of a $((2N+2)\times (2N+2))$-matrix. We  provide the explicit expressions for the energy expectation values $\braket{\hat H'}_{\mathrm{PV}}$ of the studied Hamiltonians in Appendix~\ref{exp_vals}, and detail how to efficiently compute $\mathbf H_m'$ even in cases of ill-behaved covariance matrices in Appendix~\ref{pfaffian_derivative}.

\section{Numerical results\label{num_results}}

\subsection{Non-interacting spin Hamiltonian\label{non_interacting_numerics}}
While the ground state and dynamics of the non-interacting spin-1/2 Hamiltonian of Eq.~\eqref{nis1} is exactly described by a simple spin product state, a PP-FGS Ansatz fails to do so. This is due to the fact that the resulting fermionic Hamiltonian is PV. However, it has been shown that a PV-FGS as in Eq.~\eqref{ppprime} can describe arbitrary spin-1/2 product states \cite{henderson2024hartree,lyu2024displacedfermionicgaussianstates}. We numerically demonstrate that the EOMs derived for the PV-FMFT lead to the ground state of the Hamiltonian of Eq.~\eqref{nis1}, and give an analytical explanation as to why this is expected in Appendix~\ref{product_state_construction_from_PVFGS}. While this is a proof-of-principle demonstration, it shows how PV-FMFT allows to study quenches from arbitrary initial spin-1/2 product states. 

\begin{figure}[t]
  \centering
  \subfigure{\includegraphics[width=0.9\columnwidth]{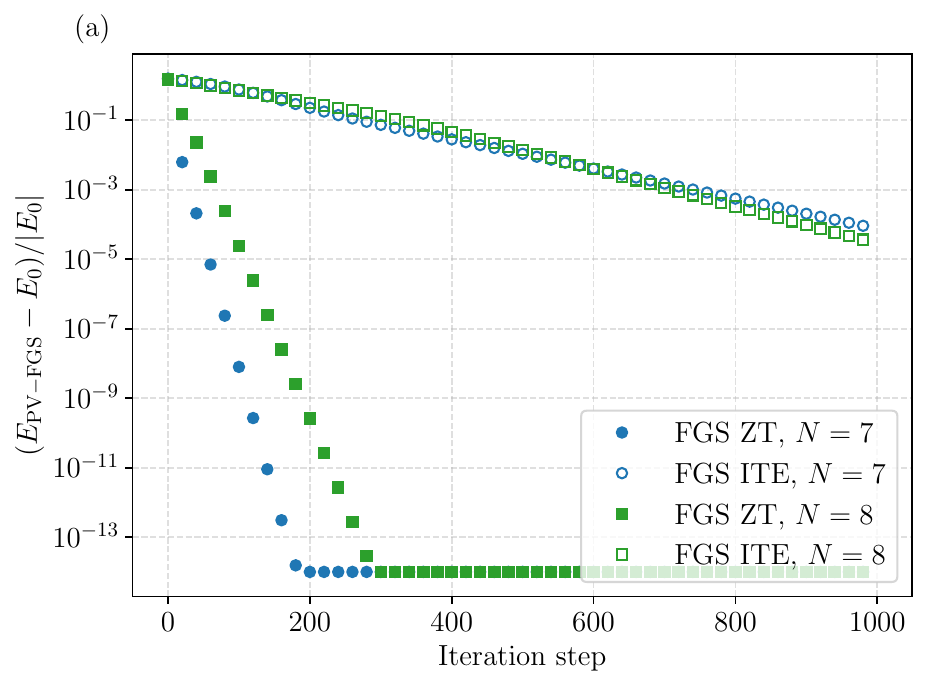}}
  % \vspace{0.5em}
  \subfigure{\includegraphics[width=0.9\columnwidth]{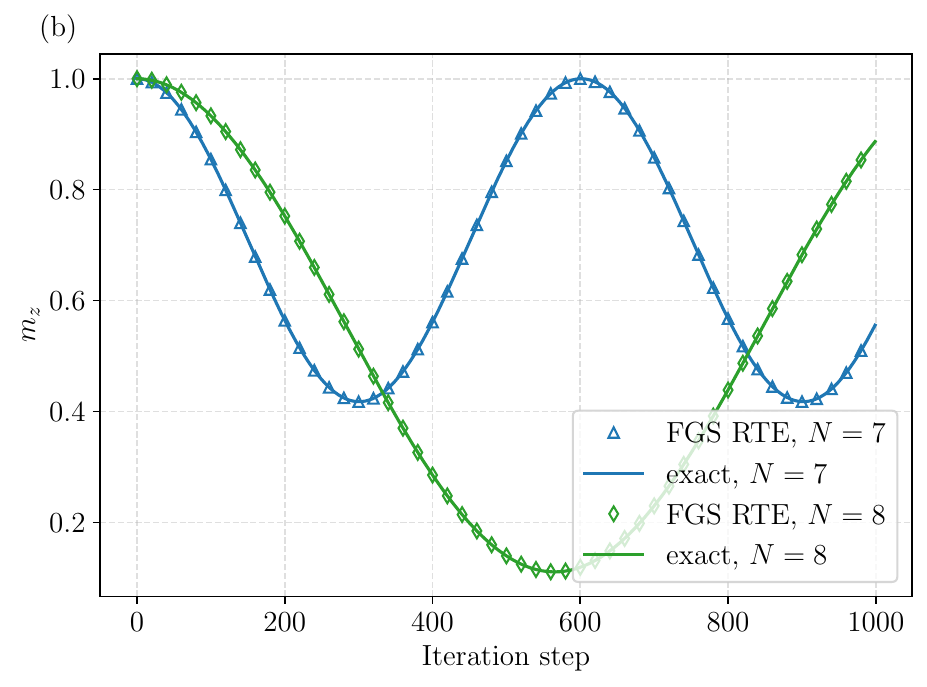}}
  \caption{Results for the PV-FMFT applied to the random non-interacting spin Hamiltonian of Eq.~\eqref{nis1} for 7 (blue) and 8 (green) spins, respectively. (a) Reaching the ground state starting from a random PV-FGS using the FGS ZT method (full markers) and the FGS ITE method (hollow markers). (b) Single site magnetization for the PV-FMFT method (markers) and the exact evolution (solid lines) after a quench from the initial state $\ket{\mathbf 0}$.}
\label{fig:noninteracting_results}
\end{figure}

Figure~\ref{fig:noninteracting_results} (a) shows the convergence of the relative energy $(E_\mathrm{PV-FGS}-E_0)/|E_0|$ of the PV-FMFT solution to the exact ground state energy $E_0$ obtained through exact diagonalization for two system sizes containing $N\in \{7,8\}$ spins respectively. In both cases, the Hamiltonian parameters $J^\alpha_p$ as well as the initial state described by $\mathbf{M}$ and $\mathbf{u}$ are randomly chosen. The plot shows the results from using two distinct numerical iterative methods. The first method (filled markers, `FGS ZT') describes the zero temperature limit $\beta\rightarrow\infty$ of an algorithm which minimizes the free energy of a given Hamiltonian within the family of FGS by solving the fixed point equation \cite{kraus2010generalized}
\begin{align}
    \boldsymbol\Gamma' = -\tan\left(\frac{\beta}{2} \mathbf H_m'(\boldsymbol\Gamma')\right).\label{is38}
\end{align}
The second numerical method (hollow markers, `FGS ITE') finds the PV-FGS which minimizes the energy by evolving a random initial covariance matrix through orthogonal rotations
\begin{align}
    \boldsymbol{\Gamma'}(\tau+\Delta\tau) =& \mathbf O_{\text{ITE}}(\tau+\Delta\tau)\boldsymbol\Gamma'(\tau)\mathbf O_{\text{ITE}}(\tau+\Delta\tau)^T,\label{d1}
\end{align}
where we defined 
\begin{align}
    \mathbf O_{\text{ITE}}(\tau+\Delta \tau) =& e^{-\frac{1}{2}\left[\boldsymbol{\Gamma}'(\tau),\mathbf H_m'(\tau)\right]\Delta \tau}.\label{d4}
\end{align}
We use a step size of $\Delta\tau=0.01$ for the FGS ITE data shown in Figure~\ref{fig:noninteracting_results} (a). For both numerical methods, PV-FMFT can reach the exact ground state energy of non-interacting spin systems. Note, that unlike the ZT method, the convergence of the ITE method depends on the step size $\Delta\tau$. 

Next, we study the real time dynamics using PP-FMFT.  Interestingly, even though the corresponding fermionic Hamiltonian is not quadratic, PV-FMFT can exactly reproduce the time evolution starting from an arbitrary spin-1/2 product state. We provide a proof that this can be expected in Appendix~\ref{SMFT_in_FMFT} by showing a one-to-one mapping from the spin-product variational parameters to the corresponding PV-FGS variational parameters, and that they follow identical EOMs. In our numerical simulations, we consider a quench of the initial state $\ket{\mathbf{0}}$\footnote{Note, that the covariance matrix $\boldsymbol{\Upsilon}$ of the initial state $\ket{\mathbf{0}}$ of the quench protocol leads to an ill-behaved covariance matrix in the enlarged Hilbert space $\mathcal H^{(N+1)}$. Such ill-behaved matrices can be treated well numerically using the method described in Appendix~\ref{pfaffian_derivative}. } with the random non-interacting Hamiltonian defined in Eq.~\eqref{nis1} and compute the time evolution of the magnetization of the first site $m_z = \braket{\hat\sigma^z_1}_{\mathrm{PV}}$ of the original spin system; The results are shown in Figure~\ref{fig:noninteracting_results} (b). The corresponding EOMs for the covariance matrix are solved using a RK4 method at a step size $\Delta t=0.01$. Figs.~\ref{fig:noninteracting_results} (a) and (b)  as well as subsequent figures only show a fraction of the data points of the iterative methods for readability. We have restricted ourselves here to small system sizes for simplicity. PV-FMFT is able to exactly capture the dynamics of the non-interacting spin system.

\subsection{Transverse-field Ising Hamiltonian\label{tfim}}
\begin{figure}[t]
  \centering
  \subfigure{\includegraphics[width=0.99\columnwidth]{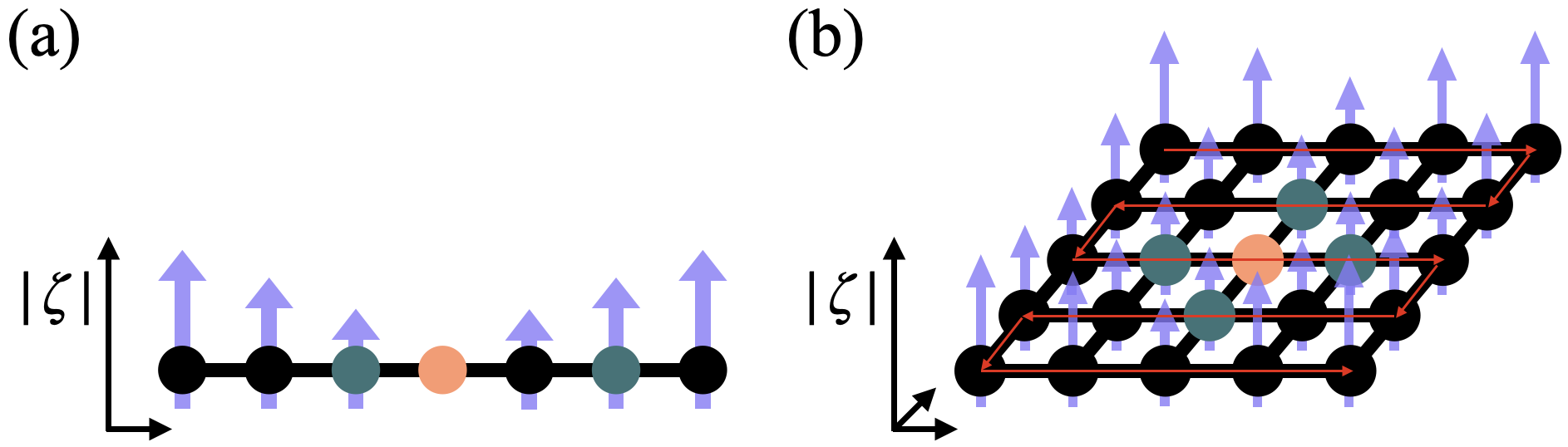}}
% \vspace{0.5em}
%   \subfigure{\includegraphics[width=0.85\columnwidth]{connected_correlations_selections.png}}
  \caption{Sketch of the longitudinal field strengths $|\zeta_k|$ on the (a) one-dimensional spin chain and (b) two-dimensional square lattice. The thin red line in the 2D square lattice indicates the snake-ordering used throughout the manuscript. The central sites (orange node) experience a zero longitudinal field $|\zeta_\mathrm{c}|=0$, while all other sites experience non-zero fields that increase with distance from the center. The green sites denote the NN sites considered for the averaged connected correlation function of the central site. For the one-dimensional chain and nearest-neighbor interactions ($\alpha\rightarrow\infty$) only the boundary sites have a non-zero longitudinal field component. }
    \label{fig:zeta_fields}
\end{figure}

We now study the TFIM with power-law interactions that can be realized in Rydberg-atom arrays with an interaction range determined by the Rydberg blockade radius~\cite{jaksch2000fast,Scholl2021}. With uniform Rabi frequency \(\Omega\) (transverse field) and detuning \(\Delta\) (longitudinal field), the Hamiltonian reads
\begin{equation}
  H=\frac{1}{2}\sum_{k\neq l} J_{kl}\,\hat n_k \hat n_l + \frac{\Omega}{2}\sum_k \hat\sigma_k^x
    - \Delta\sum_k \hat n_k,
  \label{n112}
\end{equation}
where \(\hat n_k=\tfrac{1}{2}(\mathds{1}-\hat\sigma_k^z)\) and we assume arbitrary units of energy, where $\text{max}(J_{kl})=1$.
Motivated by experimental constraints, we take \(\Omega\) and \(\Delta\) to be spatially uniform and choose $\Delta=\frac{1}{2}\sum_{l\neq c}J_{lc}$, where \(c\) denotes a central lattice site. This sets the site-dependent longitudinal field
\begin{align}
    \zeta_k \equiv \Delta-\tfrac{1}{2}\sum_{l\neq k} J_{kl}\label{zeta_chosen}
\end{align}
to zero at the center and nonzero elsewhere, with the largest magnitude near the boundaries; see Figure~\ref{fig:zeta_fields}.

We use power-law couplings \(J_{kl}=r_{kl}^{-\alpha}\) with lattice spacing set to unity, where \(r_{kl}\) is the Euclidean distance between sites \(k\) and \(l\).
Most simulations employ van der Waals interactions (\(\alpha=6\)), relevant for Rydberg platforms. Because the large exponent yields a rapidly decaying tail, we expect PV-FMFT to approximate the one-dimensional TFIM ground state well, consistent with Ref.~\cite{Kaicher2023}. In contrast, in the strong long-range regime (characterized by small \(\alpha\leq 1\)) the accuracy of PP-FMFT deteriorates; we expect a similar limitation for PV-FMFT, which we illustrate on small system sizes.

\paragraph*{Observables:} The connected correlation function between two sites $p$ and $q$ can be computed through 
\begin{align}
C_{pq}=&\braket{\hat \sigma_p^z\hat \sigma_q^z}-\braket{\hat \sigma_p^z}\braket{\hat \sigma_q^z}\nonumber\\
=&-\Gamma'_{2p,2q}\Gamma'_{2p+1,2q+1}+\Gamma'_{2p,2q+1}\Gamma'_{2p+1,2q},\label{ccf1}
\end{align}
in case of a $ZZ$-type interaction and 
\begin{align}
C_{pq} =& \braket{\hat \sigma^x_p\hat \sigma^x_q}-\braket{\hat \sigma^x_p}\braket{\hat \sigma^x_q} \nonumber\\
 =& (-1)^{q-p}\text{Pf}\left(\left.\boldsymbol\Gamma'\right|_{2p+1,\dots,2q}\right)\nonumber\\&-(-1)^{p+q}\text{Pf}\left(\left.\boldsymbol\Gamma'\right|_{1,\dots,2p}\right)\text{Pf}\left(\left.\boldsymbol\Gamma'\right|_{1,\dots,2q}\right)\label{ccf2}
\end{align}
for a $XX$-type interaction. The indices $p,q$ on the left-hand side of the equation in the first identities denote the spin indices, while in the last identities they denote the fermionic modes in the extended Hilbert space. In this work, we compute the central site magnetization $m_\mathrm{c}$ and the connected correlation function at the central site $C^{\mathrm{NN}}_\mathrm{c} = \tfrac{1}{|\{j_c\}|_{j_c\in J_c}}\sum_{j\in J}C_{c,c+j_c}$, where $J_c$ denotes the set of indices of the chosen nearest-neighours of the central site $c$, along the direction of the interaction (longitudinal) for the one- and two dimensional TFIM described by Eq.~\eqref{n112}. We consider different ways of computing $C^{\mathrm{NN}}_\mathrm{c}$ based on different nearest-neighbor sets $J_c$, see Figure~\ref{fig:2d_tfim_symmetry_results} (a). This is relevant for the two dimensional square lattice, whose indices are labeled according to the snake labeling shown in Figure~\ref{fig:zeta_fields} (b). 

\paragraph*{Influence of spin-basis representation:} 
In one dimension and with open boundary conditions (OBC), the TFIM with NN couplings and no longitudinal field is exactly solvable after a JW transformation: within the $XX$-representation
\(
H=J\sum_i\sigma_i^x\sigma_{i+1}^x + h\sum_i\sigma_i^z
\)
the JW image is a quadratic fermionic operator. Consequently, the ground state---and real-time dynamics whenever the initial state is a PP-FGS---are reproduced exactly by PP-FMFT. By contrast, the $ZZ$-representation
\(
H=J\sum_i\sigma_i^z\sigma_{i+1}^z + h\sum_i\sigma_i^x
\)
is related to the $XX$-representation by a uniform spin rotation, but under the standard JW-transformation of  Eqs.~\eqref{j2}-\eqref{j4} the fermionic interaction terms are no longer quadratic. Hence PP-FGS ceases to be exact in that basis. More generally, although the two spin Hamiltonians are unitarily equivalent, different spin bases lead---after fermionization---to different variational problems: single-spin rotations are not Gaussian transformations, so they move the model closer to or farther from the `best' fermionic-Gaussian manifold. This is why the choice of spin basis (before JW) measurably affects the quality of an FMFT approximation for interacting cases, see e.g. Refs.~\cite{henderson2023restoringpermutationalinvariancejordanwigner,henderson2024hartree}. To see the effect of the chosen spin basis, we study PV-FMFT for both the $XX$- and $ZZ$-representation for the one- and two-dimensional TFIM.

\paragraph*{Numerical PV-FMFT solver:} A direct real-time update similar to Eq.~\eqref{d1} was numerically unstable for the TFIM. We therefore integrate in time with a fourth-order Runge–Kutta (RK4) method using a fixed step size $\Delta tJ=0.001$, which let to stable results across all simulated system sizes, dimensions and parameter regimes of the TFIM.

\paragraph*{MPS details:}
The matrix product state (MPS) simulations presented in this work were performed using the \texttt{emu\_mps} Python package~\cite{Bidzhievpasqalemulators2025, bidzhiev2025efficientemulationneutralatom}, employing a two-site time-dependent variational principle (TDVP) algorithm. For one-dimensional systems, the MPS mapping follows the standard linear chain representation, while for two-dimensional systems a horizontal snaking pattern was used, mapping each row from left to right. Time evolution was carried out with a time step of $\Delta t J \approx 0.01$. To ensure numerical reliability, the bond dimension was chosen according to system size and computational constraints: for systems of 81 sites in both one and two dimensions, we utilized a bond dimension of 900, the largest available in multiples of 100, that could be supported by a 40~GB NVIDIA A100 GPU, whereas for smaller systems a fixed bond dimension of 1000 was sufficient to produce reliable dynamical properties. It is worth noting that for large systems and long-time evolutions, MPS calculations may lose quantitative accuracy when $h_x=\mathcal O(J)$ due to the rapid growth of entanglement, particularly in two-dimensional systems~\cite{vovrosh2025simulatingdynamicstwodimensionaltransversefield}.

\paragraph*{TW details:} We consider a semiclassical numerical method based on the discrete truncated Wigner transformation (TW) introduced in Ref.~\cite{schachenmayer2015}. The EOMs are solved using RK4 with step size $\Delta tJ=0.001$ and a varying number of samples.

\subsubsection{One-dimensional chain}
\begin{figure}[t]
  \centering
  \subfigure{\includegraphics[width=0.9\columnwidth]{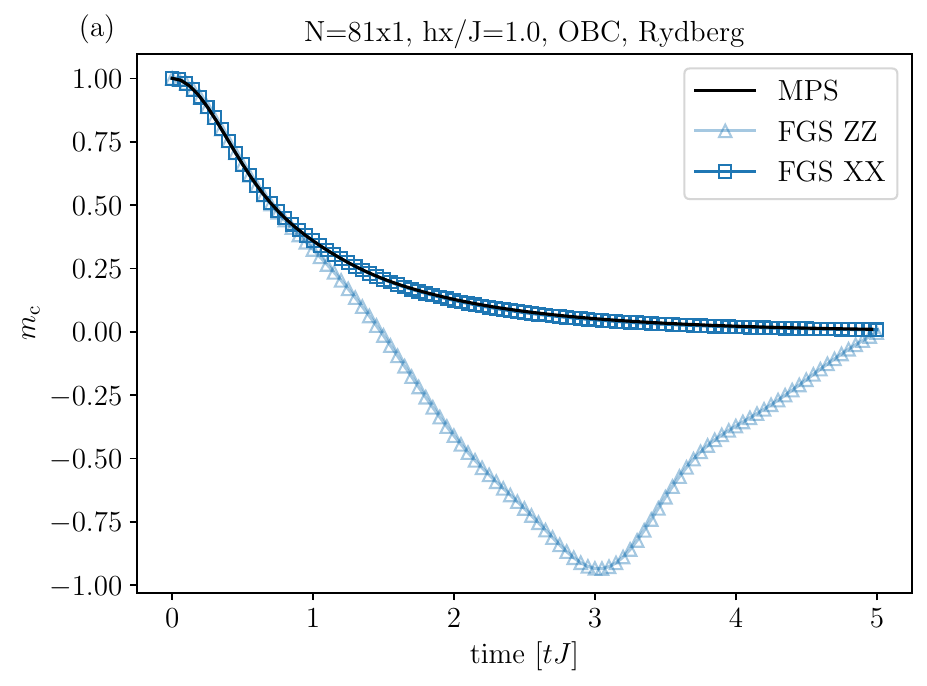}}
  % \vspace{0.5em}
  \subfigure{\includegraphics[width=0.9\columnwidth]{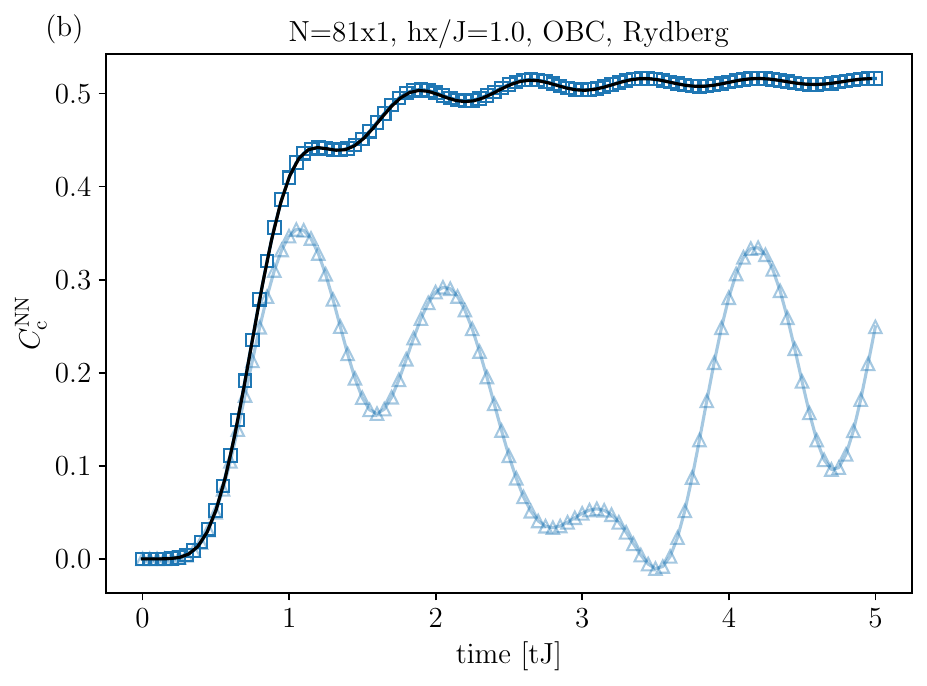}}
  \caption{(a) Single-site magnetization $m_c$  and (b) connected correlation function $C^{\mathrm{NN}}_{\mathrm{c}}$ (between c and $\mathrm{c}+1$) for the TFIM with a nonzero longitudinal field and OBC on a one-dimensional chain ($N=81$) and van der Waals interaction ($\alpha=6$) and $h_x/J=1.0$ with $h_x=\Omega/2$.  PV-FMFT ($ZZ$- and $XX$-representation) is compared against MPS results with a maximum bond dimension of $\xi=900$.}
  \label{fig:alpha_6_results_1d}
\end{figure}

We study the quench for the TFIM ($\alpha=6$) of a one-dimensional spin chain ($N=81$) with OBC and $h_x/J=1$ and a longitudinal field chosen following Eq.~\eqref{zeta_chosen}. The initial state for the quench is $\ket{\mathbf 0
}=\ket{0
}^{\otimes N}$ for the $ZZ$-representation and $\ket{\boldsymbol{+}
}=1/2^{N/2}(\ket{0
}+\ket{1
})^{\otimes N}$ for the $XX$-representation, respectively. Figure~\ref{fig:alpha_6_results_1d} shows (a) the central-site magnetization $m_c$ and (b) the nearest-neighbor connected correlator $C^{\mathrm{NN}}_c$, comparing PV-FMFT in the \(ZZ\) (light-blue triangles) and \(XX\) (dark-blue squares) representations to a converged MPS benchmark (black solid line). Both PV-FMFT curves agree with MPS at short times, but only the $XX$-representation tracks the MPS result throughout the full time window. The discrepancy is expected: under the standard JW transformation the $XX$-representation of the Ising Hamiltonian without longitudinal component becomes a quadratic (free-fermion) Hamiltonian and is exactly representable within the FGS manifold, whereas the $ZZ$-representation generates quartic fermionic terms that lie outside this manifold, where FGS provide only an approximate description. Remarkably, we observe no discrepancies between the MPS and FGS XX simulations over the timescales shown, even in the presence of a nonzero longitudinal field component and long-range tails in the interaction potential, which are expected to break integrability and should thus not be captured exactly by FGS. 

\begin{figure}[t]
  \centering
  \subfigure{\includegraphics[width=0.9\columnwidth]{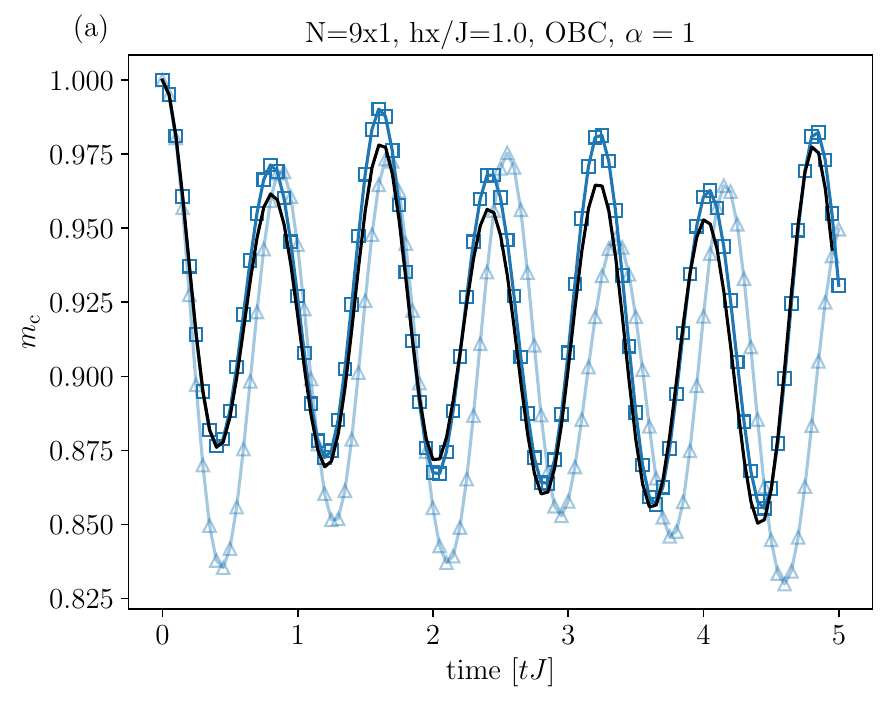}}
  % \vspace{0.5em}
  \subfigure{\includegraphics[width=0.9\columnwidth]{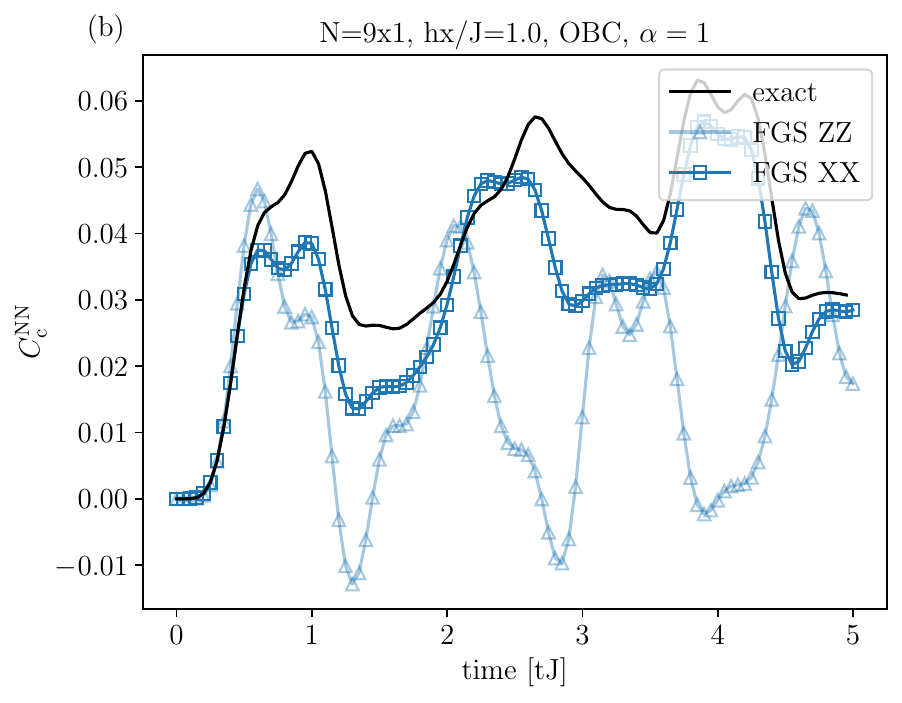}}
  \caption{(a) Single-site magnetization $m_c$  and (b) connected correlation function $C^{\mathrm{NN}}_{\mathrm{c}}$ (between c and $\mathrm{c}+1$) for the TFIM with a nonzero longitudinal field and OBC on a one-dimensional chain ($N=9$) in the strong long-range regime ($\alpha=1$) and $h_x/J=1.0$ with $h_x=\Omega/2$. PV-FMFT ($ZZ$ and $XX$ representation) is compared against exact diagonalization.}
  \label{fig:alpha_1_results_1d}
\end{figure}

The influence of the long-range tail of the interaction on the quality of the solution can be seen by moving to the strong long-range regime $\alpha=1$. In Figure~\ref{fig:alpha_1_results_1d} we study the same observables and transversal field strength, but for a smaller system size $N=9$. We compare both solutions of PV-FMFT to exact diagonalization and observe that the $XX$-representation can still qualitatively reproduce the exact solution, but no longer quantitatively. This can be explained by the increased weight of beyond-NN interaction terms (leading to non-quadratic fermionic operators) in the model.

\subsubsection{Two-dimensional square lattice}
We now turn to the two-dimensional TFIM with OBC in presence of the longitudinal field of Eq.~\eqref{zeta_chosen}. We consider the van der Waals interaction ($\alpha=6$), and a square lattice layout using the snake-ordering convention [see red arrows in Figure~\ref{fig:zeta_fields}(b)], for different values $h_x/J\in\{0.5,1.5,2.5,3.5\}$, with the value $\alpha\approx 2.5$ being in the vicinity of the dynamical phase transition~\cite{hashizume2022dynamical}.  We study the post-quench dynamics from the initial state $\ket{\mathbf 0
}=\ket{0
}^{\otimes N}$ for the $ZZ$-representation and $\ket{\boldsymbol{+}
}=1/2^{N/2}(\ket{0
}+\ket{1
})^{\otimes N}$ for the $XX$-representation, respectively, and compute the single site magnetization and connected correlation functions over time. 

We benchmark PV-FMFT against two state-of-the-art approaches: 1) matrix-product states (MPS), which provide accurate dynamics for quasi-1D geometries, and 2) the discrete truncated Wigner approximation (TW)~\cite{schachenmayer2015}, a semiclassical Monte-Carlo sampling method that has been shown to capture short-time dynamics of some large two-dimensional spin systems. The number of samples required to converge TW depends primarily on the observable rather than the system size $N$. For reference, we also include SMFT as introduced e.g. in  Ref.~\cite{mauron2024predictingtopologicalentanglemententropy}.

\begin{figure*}[!htbp]
  \centering

  \subfigure{
    \includegraphics[width=0.4\textwidth]{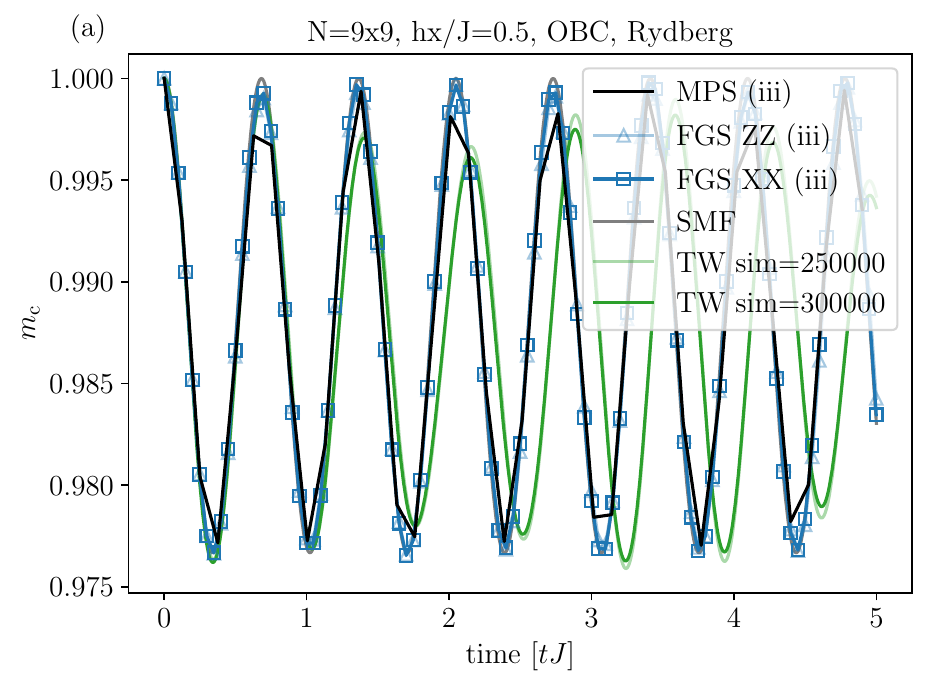}\label{fig:mgn1}
  }
  \hspace{6em} % horizontal spacing between the two figures
  \subfigure{
    \includegraphics[width=0.4\textwidth]{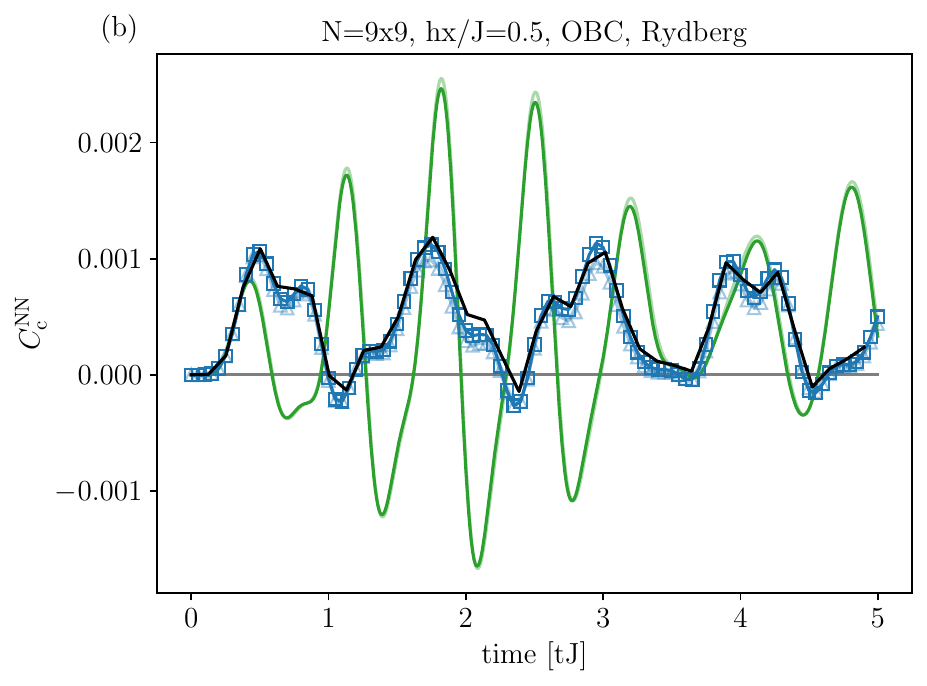}\label{fig:corr1}
  }

   \vspace{-1.5em}

   \subfigure{
    \includegraphics[width=0.4\textwidth]{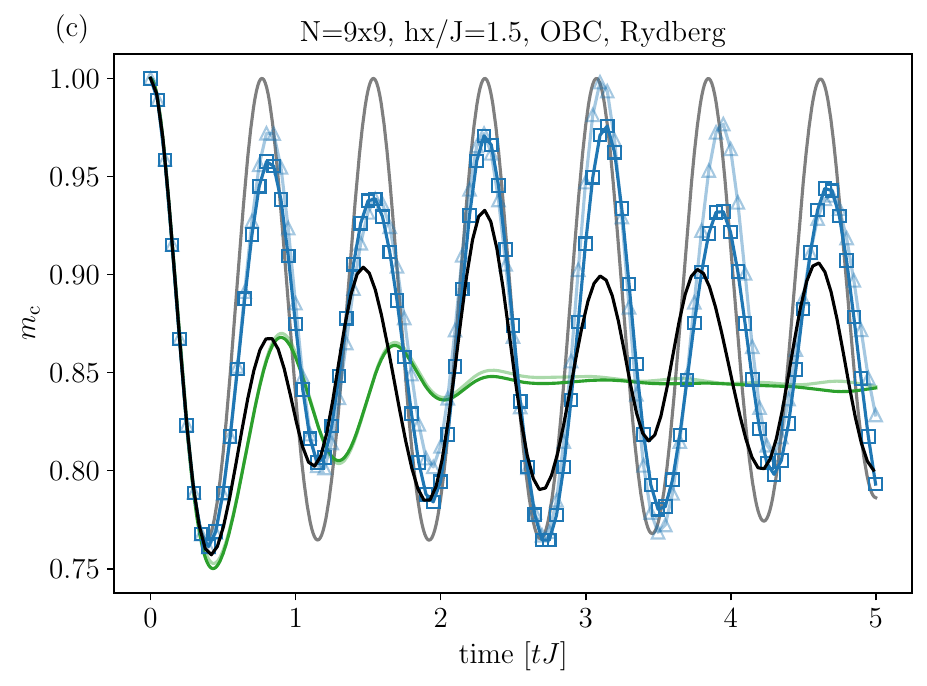}\label{fig:mgn2}
  }
  \hspace{6em} % horizontal spacing between the two figures
  \subfigure{
    \includegraphics[width=0.4\textwidth]{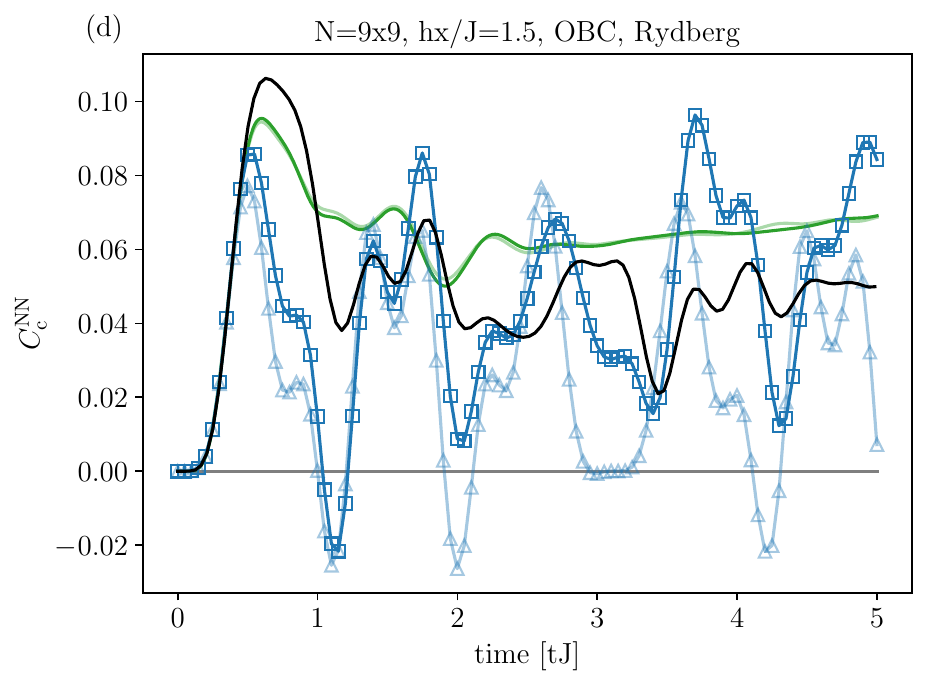}\label{fig:corr2}
  } 

   \vspace{-1.5em}

   \subfigure{
    \includegraphics[width=0.4\textwidth]{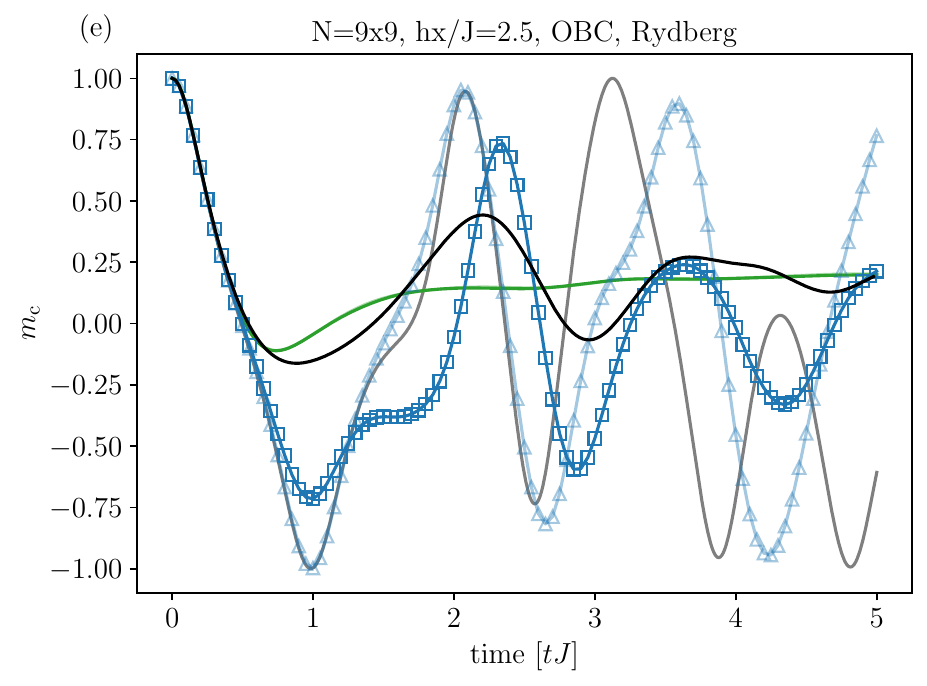}\label{fig:mgn3}
  }
  \hspace{6em} % horizontal spacing between the two figures
  \subfigure{
    \includegraphics[width=0.4\textwidth]{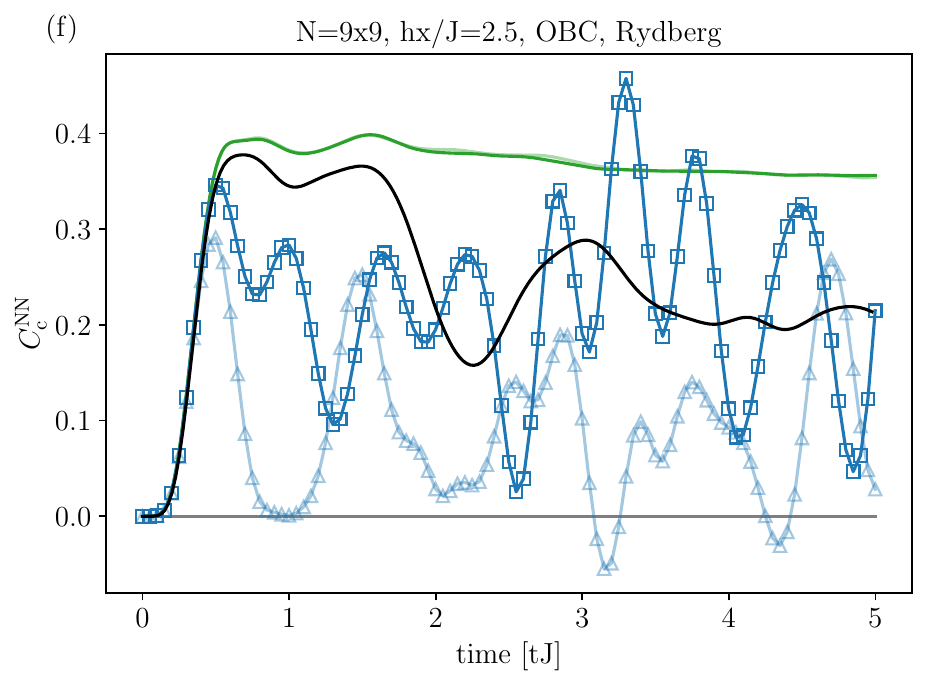}\label{fig:corr3}
  }   

   \vspace{-1.5em}

   \subfigure{
    \includegraphics[width=0.4\textwidth]{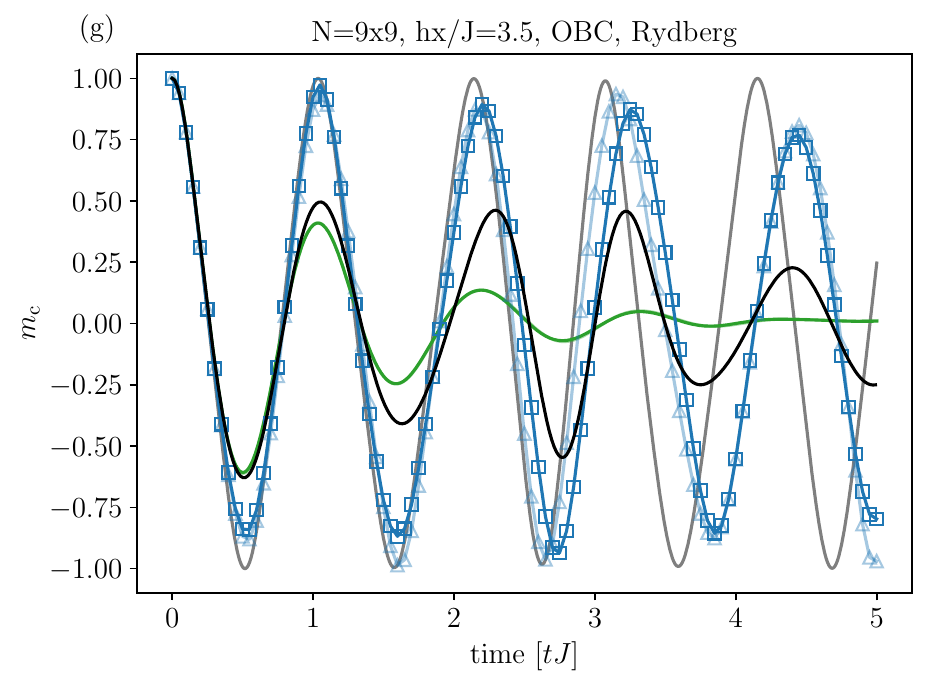}\label{fig:mgn4}
  }
  \hspace{6em} % horizontal spacing between the two figures
  \subfigure{
    \includegraphics[width=0.4\textwidth]{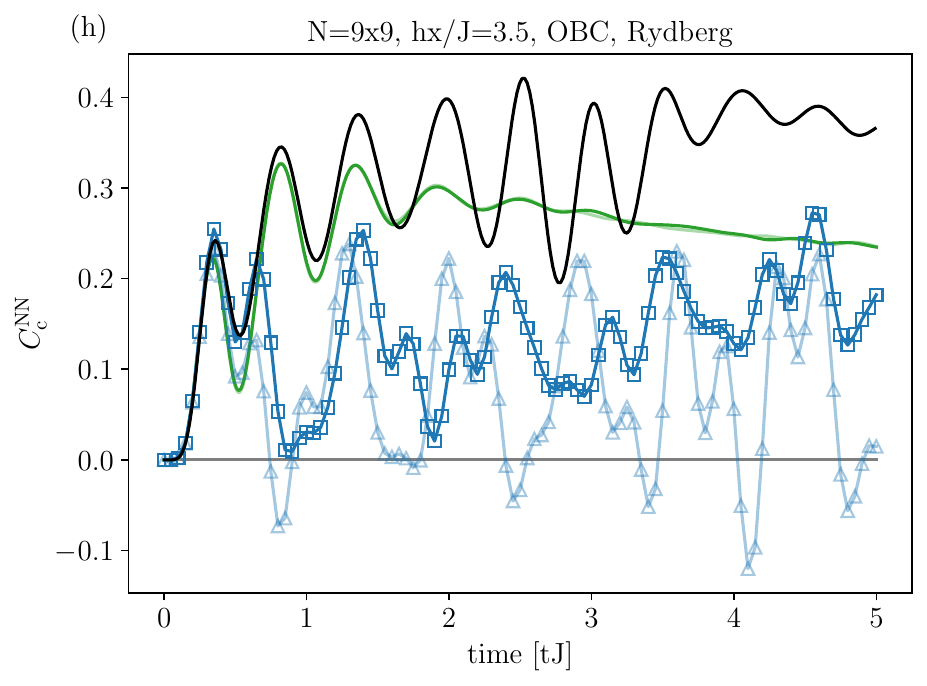}\label{fig:corr4}
  } 

  \caption{Magnetization $m_z$ at the central site (left) and averaged nearest-neighbor central connected correlation function $C^{\mathrm{NN}}_{\mathrm{center}}$ (right) of a $(9\times 9)$ square lattice TFIM with OBC and under Rydberg interaction ($\alpha =6$) for  $h_x/J\in\{0.5, 1.5, 2.5, 3.5\}$, where $h_x=\Omega/2$. The longitudinal field strengths are chosen as in Eq.~\eqref{zeta_chosen} The connected correlation functions are computed using the NN configuration as shown in Figure~\ref{fig:2d_tfim_symmetry_results} (a)(iii) for the PV-FMFT, MPS, TW and SMF. The maximum bond dimension used for the MPS simulations is $\xi=900$.}
  \label{fig:2d_tfim_results}
\end{figure*}

\begin{figure*}[!htbp]
	\centering
	
	\subfigure{
		\includegraphics[width=0.4\textwidth]{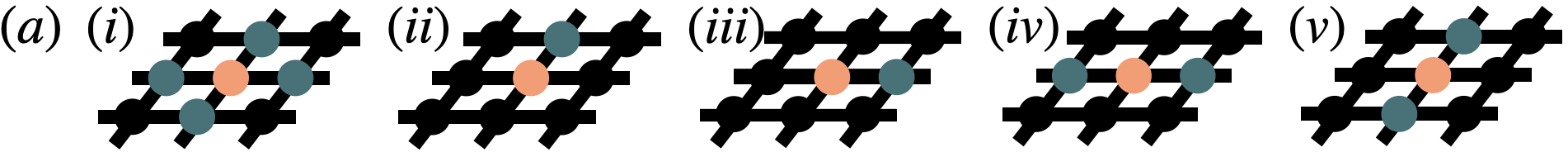}\label{fig:symmetry_labelings_2D}
	}
	
	\vspace{-1em}
	
	\subfigure{
		\includegraphics[width=0.4\textwidth]{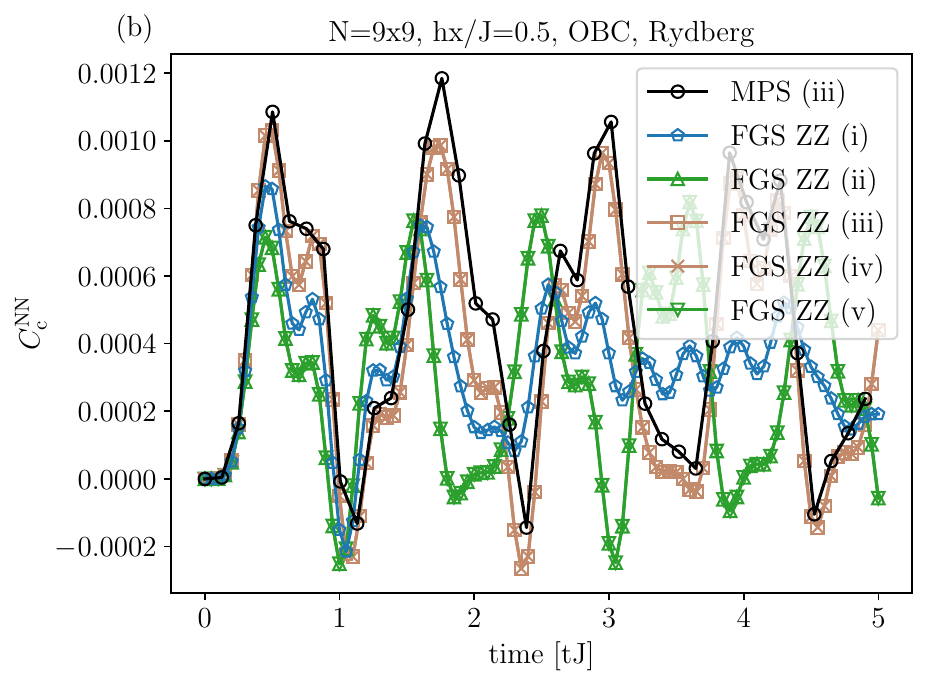}\label{fig:symm1}
	}
	\hspace{6em} % horizontal spacing between the two figures
	\subfigure{
		\includegraphics[width=0.4\textwidth]{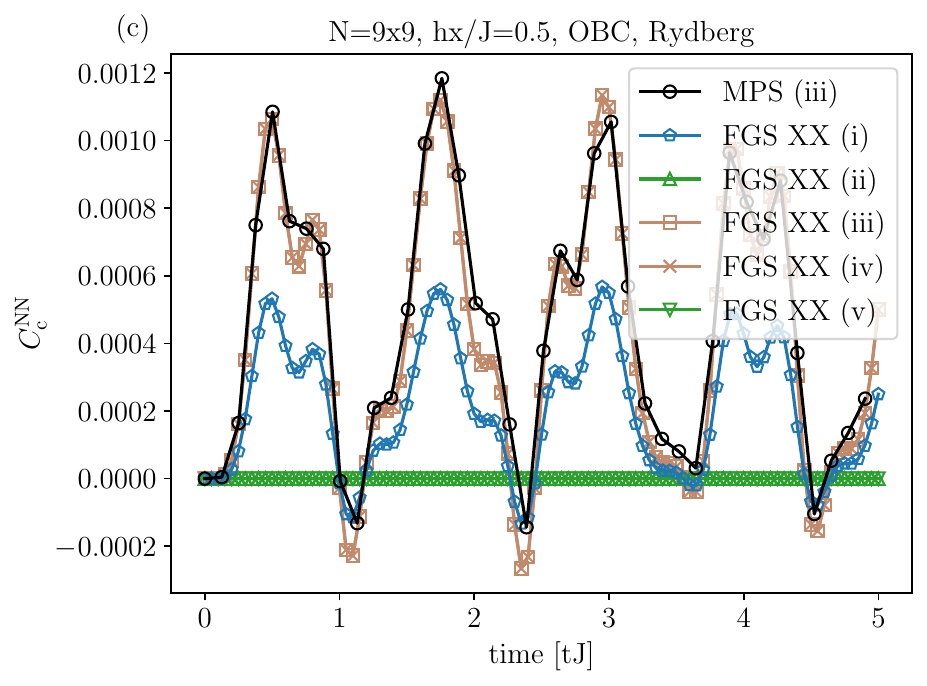}\label{fig:symm2}
	}
	
	\vspace{-1.5em}
	
	\subfigure{
		\includegraphics[width=0.4\textwidth]{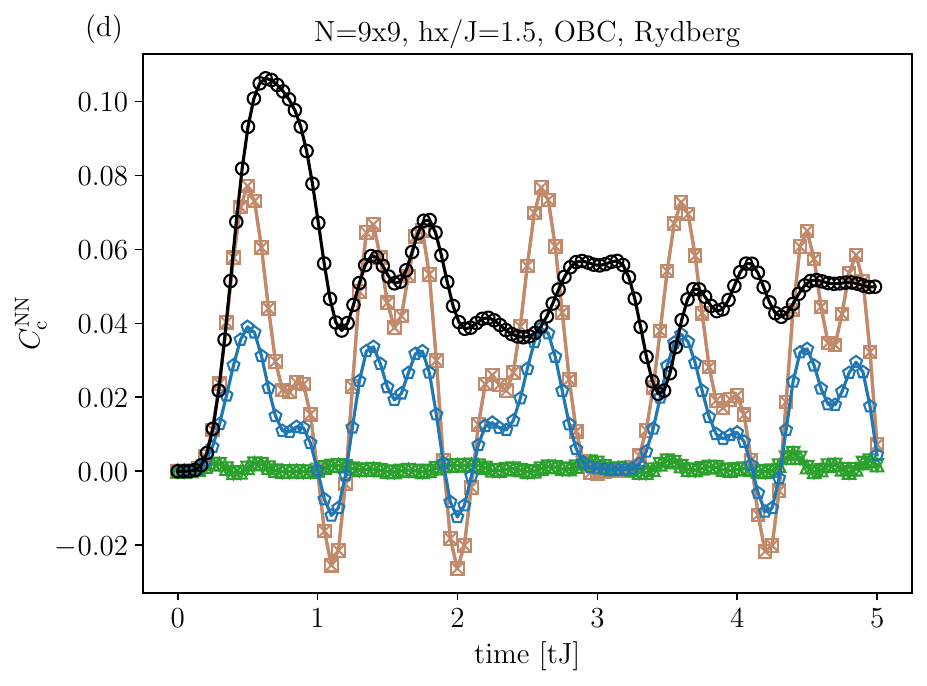}\label{fig:symm3}
	}
	\hspace{6em} % horizontal spacing between the two figures
	\subfigure{
		\includegraphics[width=0.4\textwidth]{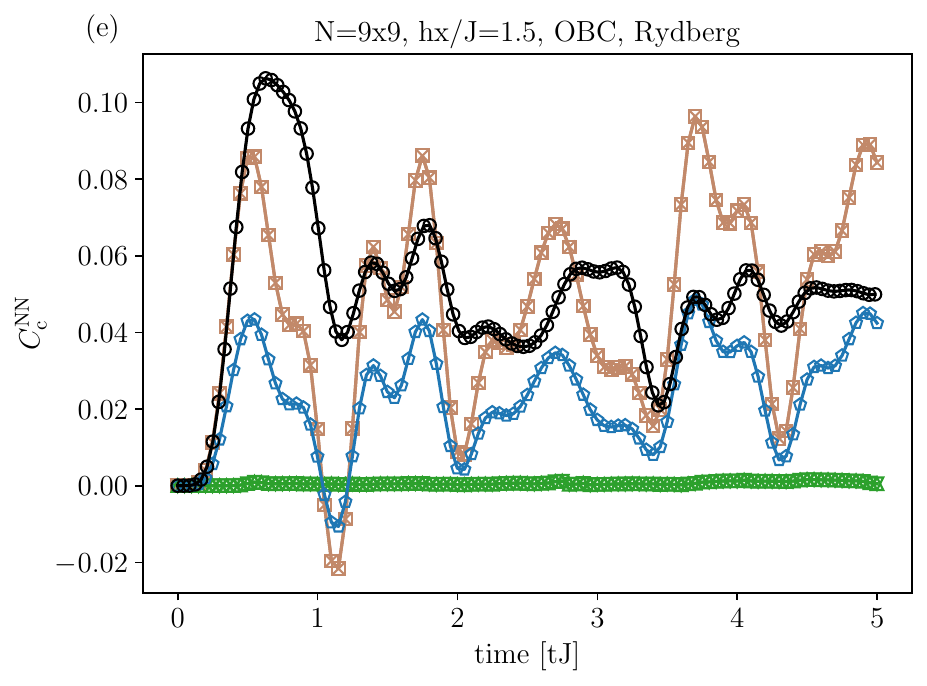}\label{fig:symm4}
	}
	
	\vspace{-1.5em}
	
	\subfigure{
		\includegraphics[width=0.4\textwidth]{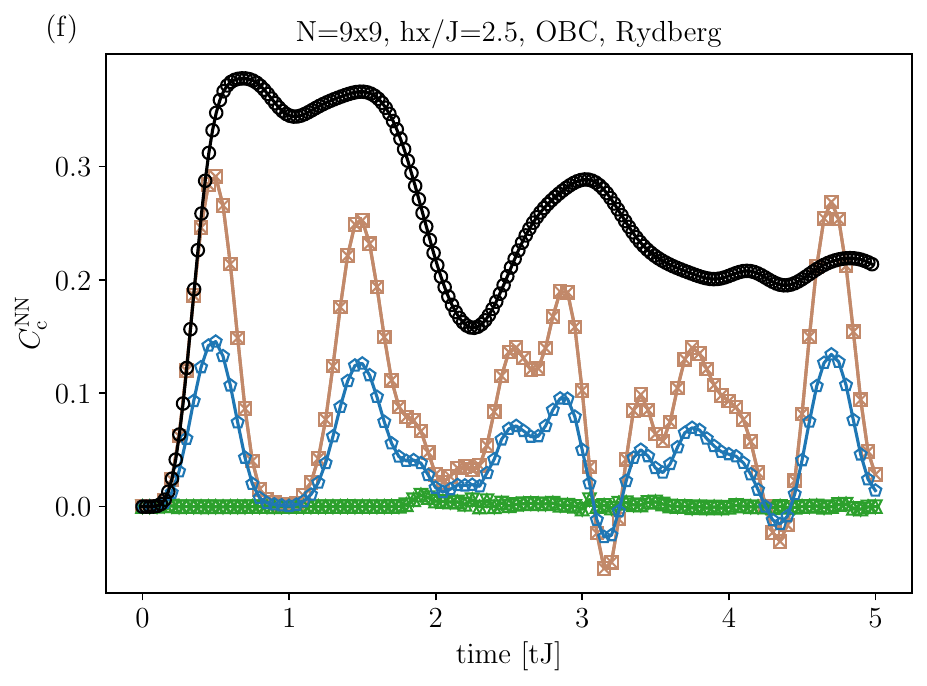}\label{fig:symm5}
	}
	\hspace{6em} % horizontal spacing between the two figures
	\subfigure{
		\includegraphics[width=0.4\textwidth]{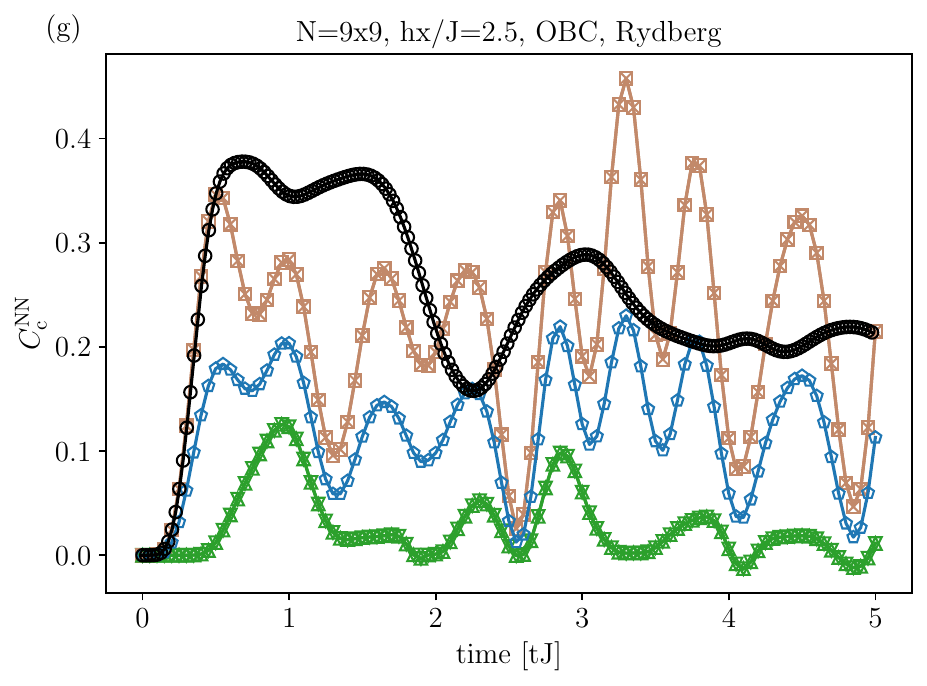}\label{fig:symm6}
	}
	
	\vspace{-1.5em}
	
	\subfigure{
		\includegraphics[width=0.4\textwidth]{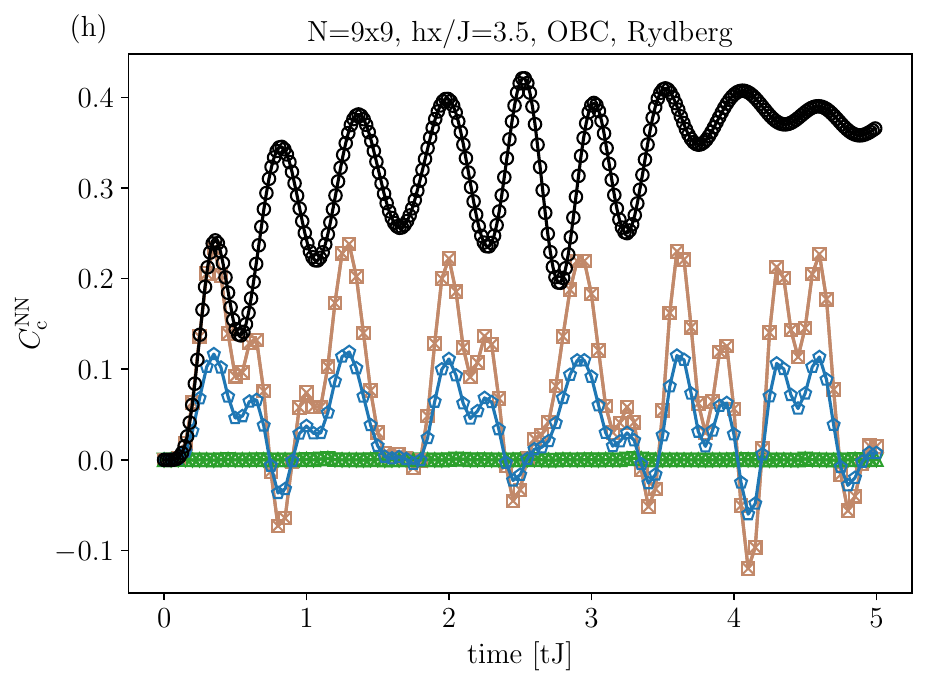}\label{fig:symm7}
	}
	\hspace{6em} % horizontal spacing between the two figures
	\subfigure{
		\includegraphics[width=0.4\textwidth]{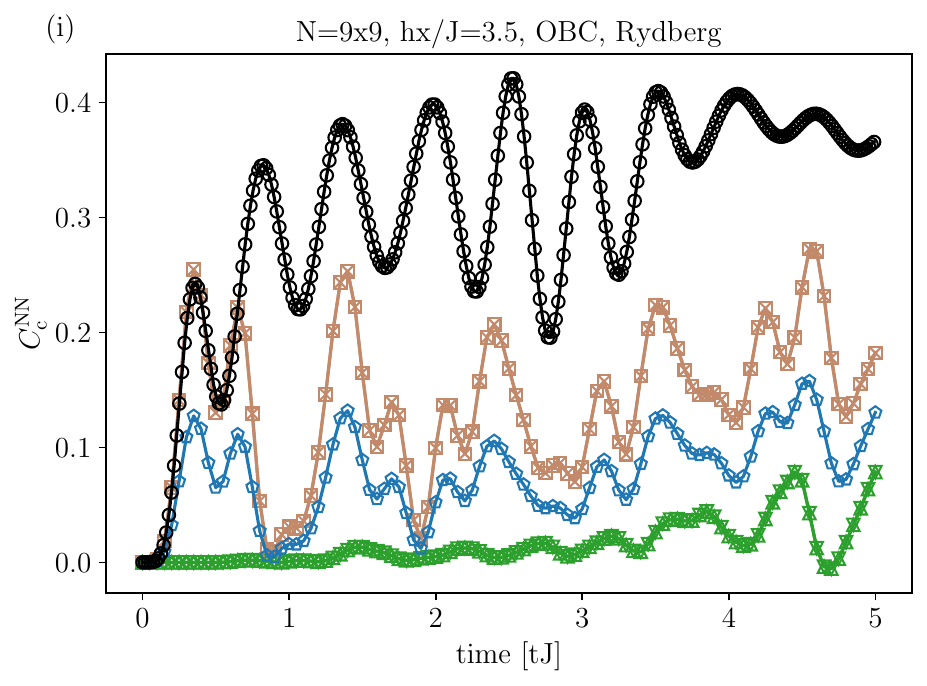}\label{fig:symm8}
	}
	
	% \vspace{-1.5em}
	
	\caption{(b)-(i): Symmetry errors for Rydberg interactions ($\alpha=6$) in connected correlation functions $C_\mathrm{c}^\mathrm{NN}$ and OBC computed using the sites displayed in the upper panel (a) (i)-(v). Left (right) column shows the results of PV-FMFT based on the TFIM in the ZZ (XX) representation and each row corresponds to a different transversal field strength. }
	\label{fig:2d_tfim_symmetry_results}
\end{figure*}

Figure~\ref{fig:2d_tfim_results} shows the post-quench dynamics of the magnetization $m_\mathrm{c}$ and connected correlation functions $C_\mathrm{c}^\mathrm{NN}$ [computed using the NN sites (iii) displayed in Figure~\ref{fig:2d_tfim_symmetry_results}) (a)] obtained from PV-FMFT in both the $ZZ$ (light blue triangles)- and $XX$-representation (dark blue squares), as well as the results from MPS (using a max. bond dimension of $\xi=900$ on a 40GB A100 GPU) and TW over two different number of samples (`sim', green lines). The SMFT solution is shown as a gray line.  In the low-transversal field regime $h_x/J=0.5$ in (a)-(b) we find that PV-FMFT reproduces both $m_\mathrm{c}$ and $C_\mathrm{c}^\mathrm{NN}$ well for both representations, while TW fails to do so. In the regime  $h_x/J=1.5$, the PV-FMFT in the $XX$-representation is still able to capture some of the qualitative feature of the dynamics. In the large-transverse field region $h_x/J=3.5$ (g)-(h) the $XX$-representation is able to capture the first oscillation of $C_\mathrm{c}^\mathrm{NN}$, but then quickly becomes unreliable. The results of TW seem to behave opposite to PV-FMFT, in that it fails to reproduce the MPS benchmark in the low-transverse field region (a)-(b), but becomes more reliable in the high-transverse field region (g)-(h) for short times before quickly converging to a classical mixture. This is consistent with the intuition that for large values $h_x/J> 1$, the system behaves more like a system of weakly interacting spins, which can typically be treated well with TW. PV-FMFT results generally improve over a simple SMF solution, but become less reliable as the transverse field strength increases. This could be traced back to the fact that the local field terms are non-quadratic and become more dominant with increasing field strengths. Both, PV-FMFT and TW fail to describe  the dynamics close to the dynamical phase transition at $h_x/J=2.5$ in (e)-(f), an indication of the fast buildup of entanglement over time. 

While the limitations of PV-FMFT in its current formulation become apparent close to the dynamical phase transition, we identify the low-transverse field region as a parameter region that can be studied for large system sizes with PV-FMFT. Overall, PV-FMFT in the $XX$-representation lead to better results than in the $ZZ$-representation. In addition, we observe that the qualitative difference in two dimensions is greatly diminished compared to the results in one dimension, which is consistent with the fact that vertical bonds (assuming our snake ordering convention) lead to non-quadratic fermionic operators in two dimensions in both representations. 

\paragraph*{Breaking of rotational symmetry: } When studying two-dimensional systems, FMFT formulated under the JW transformation breaks the rotational invariance. This symmetry breaking can be observed in the connected correlation functions in the $ZZ$- and $XX$-representations: In both cases, the transversal symmetry is broken by local magnetic fields---the transversal (longitudinal) field in the $ZZ$ ($XX$)-representation---as well as the non-local form of the interaction. 

In Figure~\ref{fig:2d_tfim_symmetry_results} we analyze the NN connected correlation function $C_{\mathrm c}^{\mathrm{NN}}$, defined in Eqs.~\eqref{ccf1}–\eqref{ccf2}, over five configurations displayed in panels Figure~\ref{fig:2d_tfim_symmetry_results}(a). Any method that preserves rotational symmetry should yield identical values across (i)–(v). With the snake ordering of Figure~\ref{fig:zeta_fields} (b),  we expect averages over horizontal bonds (iii),(iv) to differ from those over vertical bonds (ii),(v), because the latter involve longer JW parity strings. This is precisely what we find for PV-FMFT in both spin bases: within each orientation the results are consistent—(iii) agrees with (iv), and (ii) with (v)—but horizontal and vertical averages differ, demonstrating a breaking of rotational symmetry induced by the mapping rather than by the underlying spin model. We also find that in both cases, the averages over the horizontal bonds, i.e. those which include the shortest JW-strings, lead to the results closest to the MPS solution. For the $XX$-representation, across purely vertical configurations, almost no correlation is present in the low transversal field regime $h_x/J=0.5$, see Figure~\ref{fig:2d_tfim_symmetry_results} (c), and only slowly builds in (e), (f) (g) as the transversal field increases. The opposite behavior can be observed in the $ZZ$-basis, see left column of Figure~\ref{fig:2d_tfim_symmetry_results}. The configuration (i) which averages over all four NN (blue line) is not able to display the same amount of correlation as the horizontal bonds, due to the vertical bond contributions. We note, that similar breaking of rotational invariance can even be observed in strongly correlated methods based on tensor networks that rely on a mapping to Ans\"atze with dimension $<2$~\cite{vovrosh2025simulatingdynamicstwodimensionaltransversefield}.

\section{Summary and outlook\label{summary}}
We introduced a fermionic mean-field framework based on parity-violating fermionic Gaussian states (PV-FGS) and derived explicit equations of motion for both real- and imaginary-time evolution of the variational parameters; we refer to the resulting scheme as parity-violating fermionic mean-field theory (PV-FMFT). A key observation is that PV-FMFT on a system of $N$ modes can be recast as a parity-preserving (PP) FMFT for a state that is a linear combination of two FGS on an enlarged Hilbert space with one auxiliary fermionic mode. This makes FMFT applicable to arbitrary spin-1/2 models and to fermionic Hamiltonians which do not have to preserve the fermionic parity, and it naturally accommodates quenches from generic spin-1/2 product states. Our implementation is numerically stable and, in the worst case, scales as \(\mathcal{O}(N^3)\) with the number of spins/fermionic modes.

As a first validation, we numerically demonstrated that PV-FMFT yields the exact ground state for arbitrary non-interacting spin Hamiltonians and thus can represent any spin-1/2 product state. We further showed that post-quench single-site magnetization from arbitrary initial product states are also exactly captured by PV-FMFT.

We then benchmarked the one-dimensional TFIM in presence of a longitudinal field and show that for a chain of 81 spins the considered observables computed with PV-FMFT agree well with  MPS results for van der Waals ($\alpha=6$) interactions, where $\alpha$ describes the coupling strength $J_{kl}=1/r_{kl}^\alpha$. However, when considering strong long-range interactions ($\alpha=1$) and a small chain containing nine spins, we observe deviations between the PV-FMFT and results from exact diagonalization in the quench dynamics. However, the qualitative behavior of the time-evolved observables is still described well by PV-FMFT. 

In a two-dimensional square lattice containing $(9\times 9)$ spins, we study the post-quench dynamics under van der Waals interactions ($\alpha=6$) in presence of a longitudinal field and compare our results against two state-of-the-art numerical methods: MPS and a semi-classical method based on the discrete truncated Wigner approximation (TW). In the low transverse field regime $h_x/J=0.5$, we find good agreement between PV-FMFT and the MPS solution, while TW leads to qualitatively different results. However, when approaching the dynamical phase transition $h_x/J\approx 2.5$, PV-FMFT only coincides with MPS and TW for a short time. We find that TW coincides with the MPS solution over longer evolution times as the transversal field increases and the system becomes weakly coupled. 

A noteworthy limitation of the current formulation is the breaking of rotational symmetry visible in connected correlators of the JW-based PV-FMFT solutions: horizontal and vertical bonds map to different fermionic operator structures (parity strings) under a snake ordering, leading to anisotropic errors. Consistently, we find that the choice of spin representation (\(XX\) vs \(ZZ\)) significantly impacts accuracy across all TFIM settings.

Our formulation of PV-FMFT is closed in the sense that it discusses how to study arbitrary spin-1/2 Hamiltonians with PV-FGS. However, it is not optimal and we believe that the quality of the PV-FMFT solution can be improved by optimizing the chosen spin basis for representing the spin Hamiltonian.  Also the labeling of the lattice has a measurable effect for FMFT and should be investigated~\cite{henderson2023restoringpermutationalinvariancejordanwigner,henderson2024fermionic,Shi2018,Kaicher2021}. 

The breaking of rotational invariance we reported is due to the JW-transformation and is a non-physical artifact of the underlying method. Other mappings exist, which could potentially mitigate the effect of this symmetry breaking ~\cite{Verstraete2005mapping,po2021symmetricjordanwignertransformationhigher,Li2022higher,KITAEV20062}. 

It was recently shown that the PV-FGS Ansatz we employ can be implemented efficiently on digital quantum computing devices~\cite{lyu2024displacedfermionicgaussianstates} and our work thus can be employed as a means of state initialization. Another interesting  question is a direct comparison of PV-FMFT with PP-FMFT based on the PP Kitaev-mapping~\cite{KITAEV20062,rao2025dynamical,knolle2018dynamics,yilmaz2022phasediagramskitaevmodels}. Also, a comparison of PV-FMFT with their bosonic counterpart, i.e. a mean-field theory based on bosonic Gaussian states  \cite{christianen2023chemistry,Shi2018}-both in terms of expressibility and computational runtime across a range of different physical models could be an interesting topic of future research.

\section*{Acknowledgment}
We acknowledge helpful discussions with Sergi Juli\`a-Farr\'e and Tiago Mendes-Santos.  M.K. thanks Corinna Kollath for hospitality and fruitful discussions. Pasqal's team acknowledge funding from the European Union the projects PASQuanS2.1 (HORIZON-CL4-2022-QUANTUM02-SGA, Grant Agreement 101113690). S.B.J. acknowledges support
from the Deutsche Forschungsgemeinschaft (DFG, German Research Foundation) under Project No.
277625399-TRR 185 OSCAR (“Open System Control of Atomic and Photonic Matter”, B4) and under Germany’s Excellence Strategy – Cluster of Excellence Matter and Light for Quantum Computing (ML4Q) EXC 2004/1 – 390534769.

\bibliography{refs}

\newpage

\appendix

\onecolumngrid

\section{Derivation of the equations of motion for the parity-violating fermionic Gaussian state\label{proof_eom}}

In order to simplify notation, we will drop the prime notation which is used in the main text to distinguish between the original Hilbert space $\mathcal H^{(N)}$  and the extended Hilbert space $\mathcal H^{(N+1)}$. The proofs are build on Ref.~\cite{Shi2018} and require a normalized variational state $\ket{\Psi}$ as input. 

\subsection{Imaginary time evolution}
The imaginary-time evolution of a state is given by $\frac{d}{d\tau}\ket{\Psi}=-\left(H - E\right)\ket{\Psi}$ where $E=\braket{\Psi|\hat H|\Psi}$, which in our PV-FGS Ansatz of Eq.~\eqref{ppprime} can be written as
\begin{align}
	\frac{d}{d\tau} \hat U_{\text{FGS}}\ket{C(\mathbf 0)} = -\left(\hat H - E\right) U_{\text{FGS}}\ket{C(\mathbf 0)}.\label{q1}
\end{align}
Here,  $U_{\text{FGS}}$ is a PP unitary Gaussian operator  defined in Eq.~\eqref{npf_fgs_89} and $H$ is the Hamiltonian  which contains only even polynomials of Majorana operators (and does not contain $\hat A_0$) in the extended Hilbert space $\mathcal{H}^{(N+1)}$. 
In the following, we will derive EOMs for the covariance matrix $\boldsymbol{\Gamma}$ after a normal-ordering expansion and Gaussian approximation [see  Eq.~\eqref{a12}] of the left- and right-hand sides of Eq.~\eqref{q1}.

We first consider the vacuum covariance matrix 
 $\Upsilon_{pq}=-\frac{i}{2}\braket{0,\mathbf 0|[\hat A_p,\hat A_q]|0,\mathbf 0} $. In the following, we assume that $\hat U_{\text{FGS}}$ is generated by a skew-symmetric and Hermitian matrix $i\boldsymbol{\xi}$ [see Eq.~\eqref{npf_fgs_89}]. We consider a quadratic term $\hat A_p\hat A_q$, where we assume that $p<q$ and $p,q\in\{0,\dots 2N+1\}$. Using the normal-ordering expansion of Eq.~\eqref{a12}, we have $\hat A_p\hat A_q = :\hat A_p\hat A_q: +\braket{0,\mathbf 0|\hat A_p\hat A_q|0,\mathbf 0}$. Similarly, we can apply a unitary transformation through the FGS unitary generator
\begin{align}
	{U_{\text{FGS}}}^\dag\hat A_p\hat A_q U_{\text{FGS}} =& :{U_{\text{FGS}}}^\dag\hat A_p\hat A_q U_{\text{FGS}}: +\braket{0,\mathbf 0|{U_{\text{FGS}}}^\dag\hat A_p\hat A_q U_{\text{FGS}}|0,\mathbf 0}. \label{no2}
\end{align}
One can as well consider other linear unitary transformations, such as 
\begin{align}
	\tilde{U}_{\text{FGS}} =&U_{\text{FGS}}\hat A_0,\label{no15}
\end{align}
which is the transformation that generates the odd-parity sector FGS in Eq.~\eqref{ppprime}. This can be seen by realizing that $\ket{1,\mathbf 0}=\hat A_0\ket{0,\mathbf 0}$. It satisfies 
\begin{align}
	{\tilde{U}_{\text{FGS}}}{}^\dag \mathbf A \tilde U_{\text{FGS}} = \mathbf{\tilde{R}}\mathbf A,\label{no3}
\end{align}
where we defined the orthogonal matrix $\mathbf{\tilde R}$ with entries 
\begin{align}
	\tilde R_{pq} =& -R_{pq} +2\delta_{p0}R_{0q}.\label{q30:1}
\end{align}

The only part of the variational Ansatz that has a time dependence is $\boldsymbol{\xi}=\boldsymbol{\xi}(\tau)$. In order to compute the derivative of the Gaussian unitary, we will make use of the following identity which holds for any operator $\hat J(\tau)$ \cite{wilcox1967exponential},
\begin{align}
	\frac{d}{d\tau} e^{\hat J(\tau)} =& \int_0^1du e^{u\hat J(\tau)}\left(\frac{d}{d\tau} \hat J(\tau)\right)e^{(1-u)\hat J(\tau)}
	=\int_0^1du e^{(1-u)\hat J(\tau)}\left(\frac{d}{d\tau} \hat J(\tau)\right)e^{u\hat J(\tau)}\label{a5}.
\end{align}
Using the above identity, we compute the time derivative of the Gaussian unitary operator
\begin{align}
	\frac{d}{d\tau} \hat U_{\text{FGS}}
	=&\frac{i}{4} \int_0^1du e^{\tfrac{i}{4}u \hat{\mathbf A}^T \boldsymbol\xi \mathbf A}\hat {\mathbf A}^T \left(\frac{d\boldsymbol\xi}{d\tau}\right) \hat{\mathbf A} e^{\tfrac{i}{4}(1-u) \hat{\mathbf A}^T \boldsymbol\xi \hat{\mathbf A}}.\label{a6}
\end{align}
 Since in the extended Hilbert space we have 
\begin{align}
	e^{\tfrac{i}{4}u\hat{\mathbf A}^T\boldsymbol{\xi}\hat{\mathbf A}} A_j e^{-\tfrac{i}{4}u\hat{\mathbf A}^T\boldsymbol \xi \hat{\mathbf A}}
	=\left(e^{-iu\boldsymbol\xi}\mathbf A\right)_j\label{a7},
\end{align}
using Eq.~\eqref{a6}, we have
\begin{align}
	\frac{d}{d\tau} U_{\text{FGS}}
	=&\frac{1}{4}\hat{\mathbf A}^T\frac{d\mathbf R}{d\tau}\mathbf R^T \hat{\mathbf A} U_{\text{FGS}}\label{q4}\\
	\frac{d}{d\tau} \tilde U_{\text{FGS}}
	=&\frac{1}{4}\hat{\mathbf A}^T\frac{d\mathbf{\tilde R}}{d\tau}\mathbf{\tilde R}^T \hat{\mathbf A} \tilde U_{\text{FGS}}.\label{q4b}
\end{align}
We can thus rewrite the left-hand side of Eq.~\eqref{q1} as
\begin{align}
	\frac{d}{d\tau} U_{\text{FGS}}\left(\ket{0,\mathbf 0} + \ket{1,\mathbf 0}\right)
	=& U_{\text{FGS}} \left( U_{\text{FGS}}^\dag \frac{1}{4}\hat{\mathbf A}^T\frac{d\mathbf R}{d\tau}\mathbf R^T \hat{\mathbf A} U_{\text{FGS}}\right)\ket{0,\mathbf 0} + \tilde{U}_{\text{FGS}} \left( \tilde{U}_{\text{FGS}}^\dag \frac{1}{4}\hat{\mathbf A}^T\frac{d\mathbf{\tilde R}}{d\tau}\mathbf{\tilde R}^T \hat{\mathbf A} \tilde{U}_{\text{FGS}}\right)\ket{0,\mathbf 0},\label{q5}
\end{align}
where we use the definition of Eq.~\eqref{no15} and $\hat A_0^2=1$. Next, we want to make use of the normal-ordering expansions defined in Eq.~\eqref{a12}
\begin{align}
	U_{\text{FGS}}^\dag\hat \Xi U_{\text{FGS}} =& \langle 0,\mathbf 0|U_{\text{FGS}}^\dag\hat \Xi U_{\text{FGS}}|0,\mathbf 0\rangle + \frac{i}{4}:\hat{\mathbf A}^T \mathbf R^T\boldsymbol \Xi_m\mathbf R \hat{\mathbf A}: + \delta\hat \Xi,\label{q6}\\
	\tilde{U}_{\text{FGS}}^\dag\hat \Xi \tilde{U}_{\text{FGS}} =& \langle 0,\mathbf 0|\tilde{U}_{\text{FGS}}^\dag\hat \Xi \tilde{U}_{\text{FGS}}|0,\mathbf 0\rangle+ \frac{i}{4}:\hat{\mathbf A}^T \mathbf{\tilde R}^T\boldsymbol{\tilde\Xi}_m\mathbf{\tilde R} \hat{\mathbf A}: + \delta\hat \Xi,\label{q6b}
\end{align}
where 
\begin{align}
	\boldsymbol{\Gamma}=&-\mathbf R\boldsymbol{ \Upsilon}\mathbf R^T\label{q25}\\
	\boldsymbol{\tilde\Gamma}=&-\mathbf R\boldsymbol{\tilde \Upsilon}\mathbf R^T=-\mathbf{\tilde R}\boldsymbol{ \Upsilon}\mathbf{\tilde R}^T\label{q26}\\
	\boldsymbol{\Xi}_m=&4\frac{\partial\braket{0,\mathbf 0|U_{\text{FGS}}^\dag \hat \Xi U_{\text{FGS}}|0,\mathbf 0}}{\partial\boldsymbol{\Gamma}}\label{q27}\\
	\boldsymbol{\tilde\Xi}_m=&4\frac{\partial\braket{0,\mathbf 0|{\tilde U}_{\text{FGS}}^\dag \hat \Xi {\tilde U}_{\text{FGS}}|0,\mathbf 0}}{\partial\boldsymbol{\tilde\Gamma}},\label{q28}
\end{align}
 where $\tilde{\Upsilon}_{pq}=-\frac{i}{2}\braket{1,\mathbf 0|[\hat A_p,\hat A_q]|1,\mathbf 0} $.
The first term on the right-hand side of Eq.~\eqref{q5} can be written as 
\begin{align}
	U_{\text{FGS}} \left( U_{\text{FGS}}^\dag \frac{1}{4}\hat{\mathbf A}^T\frac{d\mathbf R}{d\tau}\mathbf R^T \hat{\mathbf A} U_{\text{FGS}}\right)\ket{0,\mathbf 0}\approx &U_{\text{FGS}}\left(\ket{L_0}+\ket{L_2}\right),\label{q7}
\end{align}
where 
\begin{align}
	\ket{L_0} =& \frac{i}{4}\Tr\left(\frac{d\mathbf R}{d\tau}\mathbf R^T \boldsymbol\Gamma\right)\ket{0,\mathbf 0},\label{q8}\\
	\ket{L_2} =&\frac{1}{4}:\hat{\mathbf A}^T \mathbf R^T\frac{d\mathbf R}{d\tau}\hat{\mathbf A}:\ket{0,\mathbf 0}.\label{q9}
\end{align}
Similarly, the odd parity sector terms are given by
\begin{align}
	\tilde{U}_{\text{FGS}} \left( \tilde{U}_{\text{FGS}}^\dag \frac{1}{4}\hat{\mathbf A}^T\frac{d\mathbf R}{d\tau}\mathbf R^T \hat{\mathbf A} \tilde{U}_{\text{FGS}}\right)\ket{0,\mathbf 0}
	\approx& \tilde U_{\text{FGS}}\left(\ket{\tilde L_0}+\ket{\tilde L_2}\right),\label{q10}
\end{align}
where
\begin{align}
	\ket{\tilde L_0} =& \frac{i}{4}\text{Tr}\left[\frac{d\mathbf{\tilde R}}{d\tau}\mathbf{\tilde R}^T\boldsymbol{\tilde{\Gamma}}\right]\ket{0,\mathbf 0},\label{q12}  \\
	\ket{\tilde L_2} =& \frac{1}{4}:\mathbf A^T\tilde{\mathbf R}^T\frac{d\tilde{\mathbf R}}{d\tau}\mathbf A:\ket{0,\mathbf 0}.\label{q13}
\end{align}

We now consider the right-hand side of Eq.~\eqref{q1}, 
\begin{align}
	-\left(H - E\right) U_{\text{FGS}}\left(\ket{0,\mathbf 0} + \ket{1,\mathbf 0}\right)=& -\left(H - E\right) U_{\text{FGS}}\ket{0,\mathbf 0}-\left(H - E\right) U_{\text{FGS}}  \ket{1,\mathbf 0}\nonumber\\
	=& -U_{\text{FGS}}U_{\text{FGS}}^\dag \left(H - E\right) U_{\text{FGS}}\ket{0,\mathbf 0}-\tilde{U}_{\text{FGS}}\tilde{U}_{\text{FGS}}^\dag \left(H - E\right) \tilde{U}_{\text{FGS}}  \ket{0,\mathbf 0}.\label{q14}
\end{align}
Since $H$ is an even polynomial and does not contain $\hat A_0$, we have 
\begin{align}
	E=\braket{0,\mathbf 0|U_{\text{FGS}}^\dag HU_{\text{FGS}}|0,\mathbf 0}=\braket{0,\mathbf 0|\tilde{U}_{\text{FGS}}^\dag H\tilde{U}_{\text{FGS}}|0,\mathbf 0}.\label{q15}
\end{align}
Using the normal-ordering expansion, we can write the first term in Eq.~\eqref{q14} as 
\begin{align}
	-U_{\text{FGS}}U_{\text{FGS}}^\dag \left(H - E\right) U_{\text{FGS}}\ket{0,\mathbf 0}\approx&U_{\text{FGS}}\ket{R_2},\label{q16}
\end{align}
where the constant terms cancel each other, and the quadratic contribution is given by 
\begin{align}
	\ket{R_2}=& -\frac{i}{4}:\hat{\mathbf A}^T \mathbf R^T\mathbf H_m\mathbf R \hat{\mathbf A}:\ket{0,\mathbf 0}.\label{q17}
\end{align}
We can write the second term in Eq.~\eqref{q14} as
\begin{align}
	-\tilde{U}_{\text{FGS}}\tilde{U}_{\text{FGS}}^\dag \left(H - E\right) \tilde{U}_{\text{FGS}}\ket{0,\mathbf 0}\approx&\tilde{U}_{\text{FGS}}\ket{\tilde R_2},\label{q18}
\end{align}
where the constant terms cancel each other, and the quadratic contribution is given by 
\begin{align}
	\ket{\tilde R_2}=& -\frac{i}{4}:\hat{\mathbf A}^T \mathbf{\tilde R}^T\mathbf{\tilde H}_m\mathbf{\tilde R} \hat{\mathbf A}:\ket{0,\mathbf 0},\label{q19}
\end{align}
and from Eq.~\eqref{q28} we have
\begin{align}
	\mathbf{\tilde H}_m = 4\frac{\partial \braket{0,\mathbf 0|\tilde{U}_{\text{FGS}}^\dag H \tilde{U}_{\text{FGS}}|0,\mathbf 0}}{\partial\boldsymbol{\tilde\Gamma}}.\label{q20}
\end{align}
Using the normal-ordering expansion to quadratic order, Eq.~\eqref{q1} can be written as 
\begin{align}
	U_{\text{FGS}}\left(\ket{L_0} + \hat A_0\ket{\tilde L_0} + \ket{L_2} + \hat A_0\ket{\tilde L_2}\right)=&U_{\text{FGS}}\left(\ket{R_2} + \hat A_0\ket{\tilde R_2}\right).\label{q30}
\end{align}

We will first look at the EOMs which follow from the quadratic contribution $\ket{L_2}$ and $\ket{R_2}$ in Eqs.~\eqref{q9} and \eqref{q17}. We define left- and right-eigenvectors of $\boldsymbol{\Upsilon}$, $(
		\hat c_0^\dag, \hat{\mathbf c}^\dag ,i\hat c_0^\dag,i\hat{\mathbf c}^\dag
	)\boldsymbol{\Upsilon} = -i(		\hat c_0^\dag,\hat{\mathbf c}^\dag,i\hat c_0^\dag,i\hat{\mathbf c}^\dag)$,
and $\boldsymbol{\Upsilon}(\hat c_0^\dag, \hat{\mathbf c}^\dag, i\hat c_0^\dag, i\hat{\mathbf c}^\dag)^T = i(\hat c_0^\dag, \hat{\mathbf c}^\dag, i\hat c_0^\dag, i\hat{\mathbf c}^\dag)^T$. The reason for considering these vectors is that we are only considering normal-ordered expressions acting on the vacuum and the only terms that do not vanish in such a normal-ordered expression are those terms that only consist of fermionic creation operators. With this in mind, we can write Eq.~\eqref{q17} as
\begin{align}
	\ket{R_2} =&-\frac{i}{4}:\hat{\mathbf A}^T \mathbf R^T\mathbf H_m\mathbf R \hat{\mathbf A}:\ket{0,\mathbf 0} 
	= -\frac{i}{4}:\begin{pmatrix}
		\hat c_0^\dag & \hat{\mathbf c}^\dag & i\hat c_0^\dag & i\hat{\mathbf c}^\dag
	\end{pmatrix} \mathbf R^T\mathbf H_m\mathbf R \begin{pmatrix}
		\hat c_0^\dag \\ \hat{\mathbf c}^\dag \\ i\hat c_0^\dag \\ i\hat{\mathbf c}^\dag
	\end{pmatrix}:\ket{0,\mathbf 0}\nonumber\\
	=& -\frac{i}{8}:\left(i\begin{pmatrix}
		\hat c_0^\dag & \hat{\mathbf c}^\dag & i\hat c_0^\dag & i\hat{\mathbf c}^\dag
	\end{pmatrix} \boldsymbol{\Upsilon}\mathbf R^T\mathbf H_m\mathbf R \begin{pmatrix}
		\hat c_0^\dag \\ \hat{\mathbf c}^\dag \\ i\hat c_0^\dag \\ i\hat{\mathbf c}^\dag
	\end{pmatrix}-i\begin{pmatrix}
		\hat c_0^\dag & \hat{\mathbf c}^\dag & i\hat c_0^\dag & i\hat{\mathbf c}^\dag
	\end{pmatrix} \mathbf R^T\mathbf H_m\mathbf R \boldsymbol{\Upsilon}\begin{pmatrix}
		\hat c_0^\dag \\ \hat{\mathbf c}^\dag \\ i\hat c_0^\dag \\ i\hat{\mathbf c}^\dag
	\end{pmatrix}\right):\ket{0,\mathbf 0}.
	\label{q33}
\end{align}
Similarly, we can write Eq.~\eqref{q9} as
\begin{align}
	\ket{L_2} =&\frac{1}{4}:\hat{\mathbf A}^T \mathbf R^T\frac{d\mathbf R}{d\tau}\hat{\mathbf A}:\ket{0,\mathbf 0} = \frac{1}{4}:\begin{pmatrix}
		\hat c_0^\dag &\hat{\mathbf c}^\dag&i\hat c_0^\dag & i\hat{\mathbf c}^\dag 
	\end{pmatrix}\mathbf R^T\frac{d\mathbf R}{d\tau}\begin{pmatrix}
		\hat c_0^\dag \\ \hat{\mathbf c}^\dag\\i\hat c_0^\dag \\i\hat{\mathbf c}^\dag 
	\end{pmatrix} :\ket{0,\mathbf 0}.\label{q34}
\end{align}
By demanding that $\ket{L_2}$  should be identical to $\ket{R_2}$, we arrive at the condition
\begin{align}
	:\begin{pmatrix}
		\hat c_0^\dag&\hat{\mathbf c}^\dag&i\hat c_0^\dag&i\hat{\mathbf c}^\dag 
	\end{pmatrix}\mathbf R^T\frac{d\mathbf R}{d\tau}\begin{pmatrix}
		\hat c_0^\dag\\\hat{\mathbf c}^\dag\\i\hat c_0^\dag\\i\hat{\mathbf c}^\dag 
	\end{pmatrix} :\ket{0,\mathbf 0}
	=& \frac{1}{2}:\left(\begin{pmatrix}
		\hat c_0^\dag &\hat{\mathbf c}^\dag & i\hat c_0^\dag & i\hat{\mathbf c}^\dag
	\end{pmatrix} \boldsymbol{\Upsilon}\mathbf R^T\mathbf H_m\mathbf R \begin{pmatrix}\hat c_0^\dag\\
		\hat{\mathbf c}^\dag \\i\hat c_0^\dag\\ i\hat{\mathbf c}^\dag
	\end{pmatrix}\right.\nonumber\\ & \left.-\begin{pmatrix}
		\hat c_0^\dag & \hat{\mathbf c}^\dag & i\hat c_0^\dag & i\hat{\mathbf c}^\dag
	\end{pmatrix} \mathbf R^T\mathbf H_m\mathbf R \boldsymbol{\Upsilon}\begin{pmatrix}
		\hat c_0^\dag\\ \hat{\mathbf c}^\dag \\i\hat c_0^\dag\\ i\hat{\mathbf c}^\dag
	\end{pmatrix}\right):\ket{0,\mathbf 0}\nonumber\\
	=& \frac{1}{2}:\left(-\begin{pmatrix}
		\hat c_0^\dag &\hat{\mathbf c}^\dag & i\hat c_0^\dag & i\hat{\mathbf c}^\dag
	\end{pmatrix} \mathbf R^T \boldsymbol{\Gamma}\mathbf H_m\mathbf R \begin{pmatrix}
		\hat c_0^\dag \\ \hat{\mathbf c}^\dag \\ i \hat c_0^\dag\\ i\hat{\mathbf c}^\dag
	\end{pmatrix}\right.\nonumber\\&\left.+\begin{pmatrix}
		\hat c_0^\dag & \hat{\mathbf c}^\dag & i\hat c_0^\dag & i\hat{\mathbf c}^\dag
	\end{pmatrix} \mathbf R^T\mathbf H_m \boldsymbol{\Gamma}\mathbf R\begin{pmatrix}
		\hat c_0^\dag \\ \hat{\mathbf c}^\dag \\ i\hat c_0^\dag\\ i\hat{\mathbf c}^\dag
	\end{pmatrix}\right):\ket{0,\mathbf 0}.\label{q35}
\end{align}
From Eq.~\eqref{q35}, using the orthogonality of $\mathbf R$,  it follows that
\begin{align}
	\frac{d\mathbf R}{d\tau}
	=&\frac{1}{2}\left(\mathbf H_m \boldsymbol{\Gamma}\mathbf R -  \boldsymbol{\Gamma}\mathbf H_m\mathbf R\right).\label{q36}
\end{align}
Under the assumption that $\mathbf R$ is orthogonal, using $\frac{d\mathbf R^T}{d\tau}=-\mathbf R^T\frac{d\mathbf R}{d\tau}\mathbf R^T$, we get 
\begin{align}
	\frac{d\mathbf R^T}{d\tau}=&\frac{1}{2}\left(\mathbf R^T\mathbf H_m\boldsymbol{\Gamma}-\mathbf R^T\boldsymbol{\Gamma}\mathbf H_m\right).\label{q37}
\end{align}
From Eq.~\eqref{q25}, it follows that 
\begin{align}
	\frac{d\boldsymbol{\Gamma}}{d\tau} =&-\frac{d\mathbf R}{d\tau}\boldsymbol{\Upsilon} \mathbf R^T-\mathbf R\boldsymbol{\Upsilon}\frac{d\mathbf R^T}{d\tau}, \label{q38}
\end{align}
which using Eqs.~\eqref{q36}-\eqref{q37} leads to the following EOM for the covariance matrix, 
\begin{align}
	\frac{d\boldsymbol{\Gamma}}{d\tau} 
	=&-\mathbf H_m -\boldsymbol{\Gamma}\mathbf H_m\boldsymbol{\Gamma}.\label{q39}
\end{align}
We now need to make sure that the derived EOMs which follow from the considerations of the quadratic contributions are consistent with the constant contributions. We show that this is indeed the case by using Eqs.~\eqref{q9} and Eq.~\eqref{q36} and the orthogonality of $\mathbf R$,
\begin{align}
	\ket{L_0} =& \frac{i}{4}\Tr\left(\frac{d\mathbf R}{d\tau}\mathbf R^T \boldsymbol\Gamma\right)\ket{0,\mathbf 0}= \frac{i}{8}\Tr\left(\mathbf H_m\boldsymbol{\Gamma}\mathbf R\mathbf R^T \boldsymbol\Gamma\right)\ket{0,\mathbf 0}-\frac{i}{8}\Tr\left(\boldsymbol{\Gamma}\mathbf H_m\mathbf R\mathbf R^T \boldsymbol\Gamma\right)\ket{0,\mathbf 0}=0,\label{q41}
\end{align}
where the last line follows from the cyclic property of the trace. 

Using the definition of the covariance matrix in Eq.~\eqref{q26}, we have 
% \begin{align}
% 	\tilde\Gamma_{pq}=&\frac{i}{2}\braket{0,\mathbf 0|(U_{\text{FGS}}\hat A_0)^\dag[\hat A_p, \hat A_q](U_{\text{FGS}}\hat A_0)|0,\mathbf 0} =\frac{i}{2}\sum_{kl}R_{pk}R_{ql}\braket{0,\mathbf 0|\hat A_0[\hat A_p, \hat A_q]\hat A_0|0,\mathbf 0}=\frac{i}{2}\sum_{kl}R_{pk}R_{ql}\tilde\Upsilon_{pq}\nonumber\\
% 	=&\frac{i}{2}\sum_{kl}\tilde R_{pk}\tilde R_{ql}\braket{0,\mathbf 0|[\hat A_p, \hat A_q]|0,\mathbf 0}=\frac{i}{2}\sum_{kl}\tilde R_{pk}\tilde R_{ql}\Upsilon_{pq}\label{q44}.
% \end{align}
% Thus, we have
\begin{align}
	\tilde\Gamma_{pq}=&-\left(\mathbf R\boldsymbol{\tilde \Upsilon}\mathbf R^T\right)_{pq}=-\left(\mathbf{\tilde R}\boldsymbol{\Upsilon}\mathbf{\tilde R}^T\right)_{pq}.\label{q45}
\end{align}
With Eq.~\eqref{q45}, we now study what follows from the condition $\ket{\tilde L_2}=\ket{\tilde R_2}$  using Eqs.~\eqref{q13} and \eqref{q19},
\begin{align}
	\frac{1}{4}:\mathbf A^T\tilde{\mathbf R}^T\frac{d\tilde{\mathbf R}}{d\tau}\mathbf A:\ket{0,\mathbf 0}=& -\frac{i}{4}:\hat{\mathbf A}^T \mathbf{\tilde R}^T\mathbf{\tilde H}_m\mathbf{\tilde R} \hat{\mathbf A}:\ket{0,\mathbf 0}.\label{q46}
\end{align}
Using the same strategy as before, we note that due to the normal-ordering, only the fermionic creation operator part of the Majorana operators remain, leading to 
\begin{align}
	:\begin{pmatrix}
		\hat c_0^\dag&\hat{\mathbf c}^\dag&i\hat c_0^\dag&i\hat{\mathbf c}^\dag 
	\end{pmatrix}\mathbf{\tilde R}^T\frac{d\mathbf{\tilde R}}{d\tau}\begin{pmatrix}
		\hat c_0^\dag\\\hat{\mathbf c}^\dag\\i\hat c_0^\dag\\i\hat{\mathbf c}^\dag 
	\end{pmatrix} :\ket{0,\mathbf 0}
	=& \frac{1}{2}:\left(-\begin{pmatrix}
		\hat c_0^\dag &\hat{\mathbf c}^\dag & i\hat c_0^\dag & i\hat{\mathbf c}^\dag
	\end{pmatrix} \mathbf{\tilde R}^T \boldsymbol{\tilde\Gamma}\mathbf{\tilde H}_m\mathbf{\tilde R} \begin{pmatrix}
		\hat c_0^\dag \\ \hat{\mathbf c}^\dag \\ i \hat c_0^\dag\\ i\hat{\mathbf c}^\dag
	\end{pmatrix}\right.\nonumber\\&\left.+\begin{pmatrix}
		\hat c_0^\dag & \hat{\mathbf c}^\dag & i\hat c_0^\dag & i\hat{\mathbf c}^\dag
	\end{pmatrix} \mathbf{\tilde R}^T\mathbf{\tilde H}_m \boldsymbol{\tilde \Gamma}\mathbf{\tilde R}\begin{pmatrix}
		\hat c_0^\dag \\ \hat{\mathbf c}^\dag \\ i\hat c_0^\dag\\ i\hat{\mathbf c}^\dag
	\end{pmatrix}\right):\ket{0,\mathbf 0}.\label{q48}
\end{align}
Equation~\eqref{q48} then leads to 
\begin{align}
	\frac{d\mathbf{\tilde R}}{d\tau} =&\frac{1}{2}\left(\mathbf{\tilde H}_m\boldsymbol{\tilde \Gamma}\mathbf{\tilde R}- \boldsymbol{\tilde \Gamma}\mathbf{\tilde H}_m\mathbf{\tilde R}\right).\label{q49}
\end{align}
Since $\mathbf{\tilde R}$ and $\mathbf{R}$ are directly related [see Eq.~\eqref{q30:1}] it is important to check if the differential equations for both are consistent. This is done by verifying if the EOMs derived in Eq.~\eqref{q49} are consistent with the EOMs derived in Eq.~\eqref{q36}. We will use the matrix form of Eq.~\eqref{q30:1}, 
\begin{align}
	\mathbf{\tilde R} =& - \mathbf K\mathbf R\label{q50}\\
	\mathbf K=& \begin{pmatrix} \tilde{\mathbf 1}_{N+1}&\mathbf 0_{N+1}\\
		\mathbf 0_{N+1} & {\mathbf 1}_{N+1}\end{pmatrix} ,\label{q51}
\end{align}
where we defined the $(N+1)\times (N+1)$-diagonal matrix
\begin{align}
	\tilde{\mathbf 1}_{N+1} =&\begin{pmatrix}
		-1&0&\cdots&0\\
		0&1&\cdots&0\\
		&&\ddots&\\
		0&0&\cdots&1\\
	\end{pmatrix}.\label{q57}
\end{align}
For the matrix defined in Eq.~\eqref{q51}, we have
	$\boldsymbol{\tilde\Upsilon}=\mathbf K\boldsymbol{\Upsilon}\mathbf K$, $\boldsymbol{\tilde\Gamma} = -\mathbf{\tilde R}\boldsymbol{\Upsilon}\mathbf{\tilde R}^T = -\mathbf{KR}\boldsymbol{\Upsilon}\mathbf{R^T}\mathbf K=\mathbf K\boldsymbol{\Gamma}\mathbf K$, and we write 
\begin{align}
	\frac{dE}{d\tilde{\Gamma}_{pq}} =& \sum_{kl}\frac{dE}{d{\Gamma}_{kl}}\frac{d\Gamma_{kl}}{d\tilde{\Gamma}_{pq}}, \label{q54}
\end{align}
where, using the skew-symmetry of the covariance matrix,  
\begin{align}
	\frac{d\Gamma_{kl}}{d\tilde{\Gamma}_{pq}} =& \frac{1}{2}\left(\frac{d\Gamma_{kl}}{d\tilde{\Gamma}_{pq}} - \frac{d\Gamma_{lk}}{d\tilde{\Gamma}_{pq}}\right)= \frac{1}{2}\left(\frac{d}{d\tilde{\Gamma}_{pq}}\sum_{rs}K_{kr}\tilde{\Gamma}_{rs}K_{sl} - \frac{d}{d\tilde{\Gamma}_{pq}}\sum_{rs}K_{lr}\tilde{\Gamma}_{rs}K_{sk}\right)= \frac{1}{2}\left(K_{kp}K_{ql} - K_{lp}K_{qk}\right), \label{q59}
\end{align}
which, if inserted into Eq.~\eqref{q54} and using $\mathbf H_m=4dE/d\boldsymbol{\Gamma}$, leads to 
\begin{align}
	\mathbf{\tilde H}_m =& \mathbf K\mathbf H_m\mathbf K.\label{q60}
\end{align}
Using the identity  $\mathbf K^2=\mathds 1_{2N}$, we can then use Eq.~\eqref{q50} and Eq.~\eqref{q36} to get 
\begin{align}
	\frac{d\mathbf{\tilde R}}{d\tau} =& -\mathbf K\frac{d\mathbf{R}}{d\tau}=-\frac{1}{2}\left(\mathbf K\mathbf H_m \boldsymbol{\Gamma}\mathbf R -  \mathbf K\boldsymbol{\Gamma}\mathbf H_m\mathbf R\right)=-\frac{1}{2}\left(\mathbf{\tilde H}_m\mathbf K \boldsymbol{\Gamma}\mathbf K\mathbf K\mathbf R -  \mathbf K\boldsymbol{\Gamma}\mathbf K\mathbf{\tilde H}_m\mathbf K\mathbf R\right)\nonumber\\
	=&-\frac{1}{2}\left(\mathbf{\tilde H}_m\boldsymbol{\tilde \Gamma}\mathbf K\mathbf R -  \boldsymbol{\tilde\Gamma}\mathbf{\tilde H}_m\mathbf K\mathbf R\right)=\frac{1}{2}\left(\mathbf{\tilde H}_m\boldsymbol{\tilde \Gamma}\mathbf{\tilde R} -  \boldsymbol{\tilde\Gamma}\mathbf{\tilde H}_m\mathbf{\tilde R}\right),\label{q62}
\end{align}
which is identical to Eq.~\eqref{q49}. Therefore, the EOMs of Eq.~\eqref{q36} are consistent with the ones derived in Eq.~\eqref{q49}. Let us now check whether the condition $\ket{\tilde L_0}=0$ is also satisfied. Using Eq.~\eqref{q12}, we compute
\begin{align}
	\text{Tr}\left[\frac{d\mathbf{\tilde R}}{d\tau}\mathbf{\tilde R}^T\boldsymbol{\tilde \Gamma}\right] =&\frac{1}{2}\text{Tr}\left[\left(\mathbf{\tilde H}_m\boldsymbol{\tilde \Gamma}\mathbf{\tilde R}\mathbf{\tilde R}^T\boldsymbol{\tilde \Gamma} -  \boldsymbol{\tilde\Gamma}\mathbf{\tilde H}_m\mathbf{\tilde R}\mathbf{\tilde R}^T\boldsymbol{\tilde \Gamma}\right)\right]=\frac{1}{2}\text{Tr}\left[\mathbf{\tilde H}_m\boldsymbol{\tilde \Gamma}\boldsymbol{\tilde \Gamma}\right]-\frac{1}{2}\text{Tr}\left[  \boldsymbol{\tilde\Gamma}\mathbf{\tilde H}_m\boldsymbol{\tilde \Gamma}\right]=0,\label{q63}
\end{align}
where we made use of the orthogonality of $\mathbf {\tilde R}$ and used the cyclic property of the trace. We have thus shown that the imaginary time evolution is described by Eq.~\eqref{q36} and that these EOMs are consistent with the odd-parity subspace EOMs in Eq.~\eqref{q49}, and satisfy the condition that the zeroth-order normal ordering expressions $\ket{L_0}$ and $\ket{\tilde L_0}$ disappear.

\subsection{Real-time evolution}
In order to derive the RTE EOMs, we proceed similarly. The underlying differential equation we need to solve is given by 
\begin{align}
    \frac{d}{dt}\ket{\Psi_{\text{FGS}}} = -i\hat H\ket{\Psi_{\text{FGS}}}.\label{r1}
\end{align}
We first consider the quadratic contribution. From Eqs.~\eqref{q9} and ~\eqref{q17}  the condition $\ket{L_2}=\ket{R_2}$ with
\begin{align}
    \ket{R_2}=& \frac{1}{4}:\hat{\mathbf A}^T \mathbf R^T\mathbf H_m\mathbf R \hat{\mathbf A}:\ket{0,\mathbf 0}\\
    \ket{L_2} =&\frac{1}{4}:\hat{\mathbf A}^T \mathbf R^T\frac{d\mathbf R}{d\tau}\hat{\mathbf A}:\ket{0,\mathbf 0}.
\end{align}
leads to 
\begin{align}
   \frac{d\mathbf R}{dt}=&\mathbf H_m\mathbf R + b\mathbf R\boldsymbol{\Upsilon}.\label{eomr}
\end{align}
Here, $b$ is an undetermined real-valued parameter which represents a gauge degree of freedom. This parameter has however no effect on the EOMs for the covariance matrix, which for the even parity sector are given by 
\begin{align}
   \frac{d\boldsymbol{\Gamma}}{dt}=&[\mathbf H_m, \boldsymbol{\Gamma}],\label{c6b} 
\end{align}
which follows directly from Eq.~\eqref{eomr}. The constant-terms are
\begin{align}
    \ket{L_0} =& \frac{i}{4}\Tr\left(\frac{d\mathbf R}{d\tau}\mathbf R^T \boldsymbol\Gamma\right)\ket{0,\mathbf 0},\\
    \ket{R_0} =&-iE\ket{0,\mathbf 0},
\end{align}
and we can choose $b$ to ensure that the constant term condition $\ket{L_0}=\ket{R_0}$ is met. 

We now move on to the odd parity sector and consider the cubic and linear contributions, 
\begin{align}
\ket{\tilde L_0} =& \frac{i}{4}\text{Tr}\left[\frac{d\mathbf{\tilde R}}{d\tau}\mathbf{\tilde R}^T\right]\ket{0,\mathbf 0}\\
\ket{\tilde L_2} =& \frac{1}{4}:\mathbf A^T\tilde{\mathbf R}^T\frac{d\tilde{\mathbf R}}{d\tau}\mathbf A:\ket{0,\mathbf 0},
\end{align}
and 
\begin{align}
\ket{\tilde R_0}=& -iE\ket{0,\mathbf 0} \\
    \ket{\tilde R_2}=& -\frac{i}{4}:\hat{\mathbf A}^T \mathbf{\tilde R}^T\mathbf{\tilde H}_m\mathbf{\tilde R} \hat{\mathbf A}:\ket{0,\mathbf 0}.
\end{align}
By choosing 
\begin{align}
    \frac{d\mathbf{\tilde R}}{d\tau} =& -\mathbf K\frac{d\mathbf{R}}{d\tau}=-\mathbf K\mathbf H_m\mathbf R -b\mathbf K \mathbf R\boldsymbol{\Upsilon}=-\mathbf K\mathbf H_m\mathbf K\mathbf K\mathbf R -b\mathbf K \mathbf R\boldsymbol{\Upsilon}=\mathbf{\tilde H}_m\mathbf{\tilde R} +b \mathbf{\tilde R}\boldsymbol{\Upsilon},
\end{align}
we have
\begin{align}    
\frac{d\boldsymbol{\tilde\Gamma}}{dt} =& -\frac{d\mathbf{\tilde R}}{dt}\boldsymbol{\Upsilon}\mathbf{\tilde R}^T - \mathbf{\tilde R}\boldsymbol{\Upsilon}\frac{d\mathbf{\tilde R}^T}{dt}= -\mathbf{\tilde H}_m\mathbf{\tilde R}\boldsymbol{\Upsilon}\mathbf{\tilde R}^T -b \mathbf{\tilde R}\boldsymbol{\Upsilon}\boldsymbol{\Upsilon}\mathbf{\tilde R}^T + \mathbf{\tilde R}\boldsymbol{\Upsilon}\mathbf{\tilde R}^T\mathbf{\tilde H}_m +b\mathbf{\tilde R}\boldsymbol{\Upsilon} \boldsymbol{\Upsilon}\mathbf{\tilde R}^T= [\mathbf{\tilde H}_m, \boldsymbol{\tilde \Gamma}],
\end{align}
and we can ensure that the condition on the constant (linear) contribution in the even (odd) parity sectors is satisfied by choosing the gauge degree of freedom to be
\begin{align}
    b=\frac{\text{Tr}\left[\mathbf H_m\boldsymbol{\Gamma}\right]+4E(\boldsymbol{\Gamma})}{2N}=\frac{\text{Tr}\left[\mathbf K^2\mathbf H_m\mathbf K^2\boldsymbol{\Gamma}\right]+4E(\boldsymbol{\Gamma})}{2N}=\frac{\text{Tr}\left[\mathbf{\tilde H}_m\boldsymbol{\tilde \Gamma}\right]+4E(\boldsymbol{\Gamma})}{2N}.
\end{align}

We note, that it is also possible to solve the EOMs for the variational parameters $\boldsymbol{\xi}$ directly through  
\begin{align}
    \frac{d\boldsymbol{\xi}}{dt} =& -i\left(  \mathbf H_m - b\boldsymbol{\Gamma} \right),\label{xi_eom}
\end{align}
where we used $\boldsymbol{\Gamma}=-\mathbf R\boldsymbol{\Upsilon}\mathbf R^T$.

\section{Energy expectation values and their derivatives\label{exp_vals}}
In order to implement the PV-FMFT, we need to compute the expectation value $E' = \braket{\Psi_{\text{PV}}'|\hat H'|\Psi_{\text{PV}}'}$ of the Hamiltonians in the extended Hilbert space w.r.t. the PV-FGS Ansatz of Eq.~\eqref{ppprime}, as well as their respective derivatives $\mathbf H_m$. The energy expectation values can be computed using Wick's theorem through Eq.~\eqref{wick}. The energy expectation value of the non-interacting spin Hamiltonian defined in Eq.~\eqref{nis3} is given by
\begin{align}
    E' =& \sum_{p=1}^N\left[\frac{J_p^x}{2}(-1)^{p}\text{Pf}\left(\left.\boldsymbol\Gamma'\right|_{1,2,\dots,2p-1,2p}\right)+\frac{J_p^y}{2}(-1)^{p}\text{Pf}\left(\left.\boldsymbol\Gamma'\right|_{1,2,\dots,2p-1,2p+1}\right)-\frac{J_p^z}{4}\left(\Gamma'_{2p,2p+1}-\Gamma'_{2p+1,2p}\right)\right].\label{nis8}
\end{align}
while the energy of the TFIM Hamiltonian defined in Eq.~\eqref{pb4} is given by
\begin{align}
 E'
 =& \frac{1}{4}\sum_{\substack{k<l\\k,l>0}}^N J_{kl}\text{Pf}\left(\left.\boldsymbol{\Gamma}'\right|_{2k,2k+1,2l,2l+1}\right)  + \sum_{k=1}^N\frac{\Omega_k^x}{2}(-1)^{k}\text{Pf}\left(\left.\boldsymbol\Gamma'\right|_{1,2,\dots,2k-1,2k}\right) - \sum_{k=1}^N\frac{\zeta_k}{2}\Gamma'_{2k,2k+1}.\label{pb5}
\end{align}
Similarly, the energy of the TFIM Hamiltonian defined in Eq.~\eqref{hd1} where interactions are of the form $XX$ is given by
\begin{align}
E' =&  \frac{1}{4}\sum_{\substack{p< q\\p,q>0}}^N J_{pq}(-1)^{q-p}\text{Pf}\left(\left.\boldsymbol{\Gamma}'\right|_{2p+1,2p+2,\dots,2q}\right)  - \frac{1}{2}\sum_{p=1}^N \Omega_p \Gamma_{2p,2p+1} +\frac{1}{2}\sum_{p=1}^N\zeta_p(-1)^{p}\text{Pf}\left(\left.\boldsymbol\Gamma'\right|_{1,2,\dots,2p-1,2p}\right).\label{hd2}
\end{align}
The expressions for $\mathbf{H}_m'=4\frac{d\braket{H'}_{\text{PV}}}{d\boldsymbol{\Gamma'}}$ follow from analytical formulas for the derivative of a Pfaffian. 

The computation of the Pfaffian and its derivative is the most costly operation in FMFT, in particular when dealing with ill-behaved (non-invertible) submatrices $\boldsymbol{\Gamma}|_I$, which for instance appear naturally when considering quenches of the TFIM model with spins initialized along the $z$-direction.

\section{Computing the gradient of the Pfaffian of ill-behaved matrices\label{pfaffian_derivative}}
We will consider two ways for computing the derivative of the Pfaffian of a ($2n\times 2n$) skew-symmetric matrix $\mathbf A$ which depends on some variables $x_j$. If the matrix $\mathbf A$ is singular for some variable $x_j$, the derivative of the Pfaffian can be computed through
\begin{align}
    \frac{\partial \text{Pf}(\mathbf A)}{\partial x_k} =& \sum_{1\leq i<j\leq 2n}(-1)^{i+j+1}\frac{\partial A_{ij}}{\partial x_k}\text{Pf}\left(\mathbf A|_{\hat i\hat j}\right),\label{grad1}
\end{align}
where we assume $i<j$ and where $\mathbf A|_{\hat i\hat j}$ is the submatrix of $\mathbf A$ with rows $i,j$ and columns $i,j$ removed. We stress, that Eq.~\eqref{grad1} can be applied also to special cases, where e.g. entire rows of a matrix $\mathbf A$ are zero, but it is numerically quite expensive since it involves having to compute $\mathcal O(n^2)$ Pfaffians of size $(2(n-1)\times 2(n-1))$. 

If $\mathbf A$ on the other hand is not singular w.r.t. $x_j$, one can use the following identity
\begin{align}
    \frac{1}{\text{Pf}(\mathbf A)}\frac{\partial \text{Pf}(\mathbf A)}{\partial x_k} =&\frac{1}{2}\text{Tr}\left[\mathbf A^{-1}\frac{\partial \mathbf A}{\partial x_k}\right],\label{grad2}
\end{align}
which allows one to compute the all entries of the gradient in a single step provided one has a closed formula for $\frac{\partial \mathbf A}{\partial x_k}$. If we consider the special case $x_k=A_{lm}$, we can write Eq.~\eqref{grad2} as 
\begin{align}
    \frac{1}{\text{Pf}(\mathbf A)}\frac{\partial \text{Pf}(\mathbf A)}{\partial A_{lm}} =&\frac{1}{2}\text{Tr}\left[\mathbf A^{-1}\frac{\partial \mathbf A}{\partial A_{lm}}\right].\label{prop0}
\end{align}
The derivative of the Pfaffian is linked to the adjugate matrix. The adjugate matrix is the transpose of the cofactor matrix $\boldsymbol{\Delta}(\mathbf A)^T=\text{adj}(\mathbf A)$, 
which is defined as \cite{Ishikawa01081995}
\begin{align}
    \Delta(\mathbf A)_{ij} = (-1)^{i+j+1+\theta(i-j)}\text{Pf}\left(\mathbf A|_{\hat i\hat j}\right)\text{Pf}(\mathbf A).
\end{align}
 The adjugate matrix satisfies the relation $\mathbf A\text{adj}(\mathbf A) = \text{adj}(\mathbf A)\mathbf A = \text{det}(\mathbf A)\mathds 1$, which in case the matrix $\mathbf A$ is well behaved relates the matrix inverse with the adjugate, 
\begin{align}
    \text{adj}(\mathbf A) = \text{det}(\mathbf A)\mathbf A^{-1}.\label{prop2}
\end{align}
In the context of the adjugate matrix, it was shown that $\text{adj}(\mathbf A)$ can be accurately computed from the inverse $\mathbf A^{-1}$, even when the ladder has been inaccurately computed~\cite{STEWART1998151}. Let us at first assume that the matrix $\mathbf A$ is well-behaved and thus its inverse exists. Then, we can use Eqs.~\eqref{prop0} and ~\eqref{prop2} to compute 
\begin{align}
    \frac{\partial \text{Pf}(\mathbf A)}{\partial A_{lm}} =&\frac{1}{2\text{Pf}(\mathbf A)}\text{Tr}\left[\text{adj}(\mathbf A)\frac{\partial \mathbf A}{\partial A_{lm}}\right]=\frac{1}{2\text{Pf}(\mathbf A)}\text{adj}(\mathbf A)_{ml}=\frac{1}{2\text{Pf}(\mathbf A)}\Delta(\mathbf A)_{lm}=\frac{1}{2} (-1)^{l+m+1+\theta(l-m)}\text{Pf}\left(\mathbf A|_{\hat l\hat m}\right),\label{prop4}
\end{align}
where we made use of the fact that we are taking the derivative of a structured skew-symmetric matrix, where $\frac{\partial \mathbf A}{\partial A_{lm}}=\mathbf J^{lm}-\mathbf J^{ml}$, where $\mathbf J^{lm}$ is the matrix which has zeros everywhere except for a 1 at row $l$ and column $m$. 

In order to compute the derivative of an ill-behaved matrix, following Ref.~\cite{STEWART1998151} we use the Schur decomposition
\begin{align}
    \mathbf A = \mathbf Q\mathbf T\mathbf Q^T,\label{prop5}
\end{align}
where $\mathbf Q$ is a real orthogonal matrix and $\mathbf T$ has the block diagonal form
\begin{align}
    \mathbf T = \begin{pmatrix}
        0 & t_1 &\\
        -t_1 & 0 & \\ 
        && 0 & t_2 &\\
        &&-t_2 & 0 & \\ 
        &&&&\ddots\\
        &&&&&0 & t_n &\\
        &&&&&-t_n & 0 &
    \end{pmatrix},\label{prop6}
\end{align}
where $t_k\in \mathds R$ are the imaginary values of the purely imaginary eigenvalue pairs of $\mathbf A$. When at least one of those pairs is very close (or identical) to zero, the matrix inverse  $\mathbf A^{-1}$ cannot be computed. However, the adjoint (and therefore the derivatives of the Pfaffian) can still be accurately computed when using an inaccurately computed matrix inverse $\mathbf A_\varepsilon$, where 
\begin{align}
\mathbf A_\varepsilon =&\mathbf Q\mathbf T_\varepsilon\mathbf Q^T,\label{prop7}
\end{align}
with 
\begin{align}
      \mathbf T_\varepsilon = \begin{pmatrix}
        0 & t_1+\varepsilon &\\
        -(t_1+\varepsilon) & 0 & \\ 
        && 0 & t_2+\varepsilon &\\
        &&-(t_2+\varepsilon) & 0 & \\ 
        &&&&\ddots\\
        &&&&&0 & t_n+\varepsilon &\\
        &&&&&-(t_n+\varepsilon) & 0 &
    \end{pmatrix},\label{prop8}  
\end{align}
and the derivative of the Pfaffian can be computed to an accuracy $\mathcal O(\varepsilon)$. Numerically, we are then able to compute the entire gradient in parallel, avoiding $\mathcal O(n^2)$ computations of the individual sub-Pfaffians $\text{Pf}(\mathbf A|_{\hat l\hat m})$ through Eq.~\eqref{prop4}. This is achieved by computing the derivative of the Pfaffian through the approximate matrix inverse of Eq.~\eqref{prop7},
\begin{align}
\frac{\partial \text{Pf}(\mathbf A)}{\partial \mathbf A} \approx&-\frac{1}{2}\text{Pf}(\mathbf A_\varepsilon)\mathbf A_\varepsilon^{-1}=-\frac{1}{2}\text{det}(\mathbf Q)\left(\prod_{k=1}^n(t_k+\varepsilon)\right)\mathbf Q\mathbf T_\varepsilon^{-1}\mathbf Q^T,\label{prop10}
\end{align}
where we used the Pfaffian properties $\text{Pf}(\mathbf{Q}\mathbf T_\varepsilon\mathbf{Q}^T)=\text{det}(\mathbf Q)\text{Pf}(\mathbf T_\varepsilon)$ and $\text{Pf}\begin{psmallmatrix}
    0&a\\-a&0
\end{psmallmatrix}=a$.

Note, that in general we will be dealing with expressions of the form $\frac{\partial \text{Pf}(\mathbf A|_I)}{\partial \mathbf A}$, where $I=\{i_1,i_2,\dots, i_{2k}\}$ with $k\leq n$ denotes some subindex set. In that case, Eq.~\eqref{prop10} can still be used, and the resulting $(2k\times 2k)$ matrix needs to be embedded into the $(2n\times 2n)$ matrix (all entries which contains rows or columns that are not in $I$ are set to zero).

\section{Arbitrary spin-1/2 product states lie in the family of PV-FGS}
\label{product_state_construction_from_PVFGS}

Following the discussion in Ref.~\cite{henderson2024hartree}, we consider a general product state of two spin-$1/2$ particles,
\begin{equation}
    \ket{\Psi}
    = (1+\eta_1\sigma_1^+)(1+\eta_2\sigma_2^+)\ket{00},
    \qquad 
    \eta_k\in\mathds{C},
\end{equation}
where $\ket{0}$ denotes a spin-down state.  Under the JW transformation, $\sigma_k^+=\hat c_k^\dagger \hat S_k$, with the string operator $\hat S_k$ defined in Eq.~\eqref{string}, the corresponding fermionic state is
\begin{equation}
    \ket{\Phi}
    = (1+\xi_1 \hat c_1^\dagger)(1+\xi_2 \hat c_2^\dagger \hat S_{1,1})\ket{\mathbf 0}
    = (1+\xi_1 \hat c_1^\dagger)(1+\xi_2 \hat c_2^\dagger)\ket{\mathbf 0},
\end{equation}
since the vacuum $\ket{\mathbf 0}$ has even fermionic parity.  Due to the anti-commutation relations of fermionic operators, we may rewrite this state as a fermionic Gaussian operator acting on the vacuum:
\begin{equation}
\begin{aligned}
    \ket{\Phi}
    &= e^{\xi_1\hat c_1^\dagger}
       e^{\xi_2\hat c_2^\dagger}
       \ket{\mathbf 0} = e^{\xi_1\hat c_1^\dagger
         +\xi_2\hat c_2^\dagger
         +\xi_1\eta_2 \hat c_1^\dagger \hat c_2^\dagger}\ket{\mathbf 0},
\end{aligned}
\end{equation}
where we made use of the Baker-Campbell-Hausdorff formula. The construction extends directly to three spins: The state
\begin{equation}
    \ket{\Psi}
    = (1+\eta_1\sigma_1^+)(1+\eta_2\sigma_2^+)(1+\eta_3\sigma_3^+)\ket{000}\label{pr2}
\end{equation}
maps under the JW transformation to
\begin{equation}
\begin{aligned}
    \ket{\Phi}
    &= e^{\xi_1\hat c_1^\dagger
         +\xi_2\hat c_2^\dagger 
         +\xi_1\xi_2\hat c_1^\dagger \hat c_2^\dagger}
       e^{\xi_3\hat c_3^\dagger}
       \ket{\mathbf 0} 
  = e^{\xi_1\hat c_1^\dagger
         +\xi_2\hat c_2^\dagger 
         +\xi_3\hat c_3^\dagger
         +\xi_1\xi_2\hat c_1^\dagger\hat c_2^\dagger
         +\xi_1\xi_3\hat c_1^\dagger\hat c_3^\dagger
         +\xi_2\xi_3\hat c_2^\dagger\hat c_3^\dagger}
         \ket{\mathbf 0}.
\end{aligned}
\end{equation}
The pattern for larger number of spins follows from the above considerations.  Thus every spin-1/2 product state can be constructed from a PV fermionic Gaussian operator acting on the fermionic vacuum state.

The resulting state is not necessarily normalized, but normalization can always be restored, as done for example in imaginary-time evolution (ITE) algorithms.  
Importantly, the appearance of linear terms in the exponent is precisely what distinguishes PV-FGS from PP-FGS: Since PP-FGS lack linear fermionic terms, they cannot represent generic spin-$1/2$ product states.  
In contrast, the Gaussian-filter Ansatz used in our ITE formulation (Sec.~\ref{theory}) naturally includes these PV contributions, and the corresponding PV-FGS will be obtained by solving the EOM in Eq.~\eqref{eom_ite}.

\section{PV-FMFT can reproduce SMFT}
\label{SMFT_in_FMFT}
We want to show that the dynamics of a system of $N$ non-interacting spin-1/2 Hamiltonians can be described by PV-FMFT. The real-time EOMs for a variational wave function $\ket{\Psi(\boldsymbol{\theta})}$ derived in Section~\ref{theory} were derived from the time dependent variational principle~\cite{Shi2018}
\begin{align}
    d_t\theta_j =&-i\sum_k(\mathbf G^{-1})_{jk}\braket{\Psi_k|\hat H|\Psi},\label{pr1}
\end{align}
where $\ket{\Psi_k}=\tfrac{\partial}{\partial\theta_k}\ket{\Psi}$ describes a tangent vector. In order to show that FMFT can exactly capture the dynamics of SMFT, we show that the dynamics resulting from Eq.~\eqref{pr1} are identical for any given spin-1/2 product state wave function. 

Following Appendix~\ref{product_state_construction_from_PVFGS}, we will consider the case $N=3$, where the general spin wave function is described by
\begin{equation}
    \ket{\Psi(\boldsymbol{\eta})}
    = (1+\eta_1\sigma_1^+)(1+\eta_2\sigma_2^+)(1+\eta_3\sigma_3^+)\ket{000}\label{pr4}
\end{equation}
In order to not confuse the variational parameters in spin and fermionic representation, we will write the fermionic wave function that represents Eq.~\eqref{pr2} as 
\begin{equation}
    \ket{\Phi(\boldsymbol{\xi})}
    = (1+\xi_1 \hat c_1^\dagger)(1+\xi_2 \hat c_2^\dagger )(1+\xi_3 \hat c_3^\dagger )
    \ket{\mathbf 0},\label{pr5}
\end{equation}
where $\xi_k\in\mathds C$. The tangent vectors are given by e.g. $\ket{\Psi_1}=\sigma_1^+(1+\eta_2\sigma_2^+)(1+\eta_3\sigma_3^+)\ket{000}$, and e.g. $\ket{\Phi_2}
    = (1+\xi_1 \hat c_1^\dagger) \hat c_2^\dagger(1+\xi_3 \hat c_3^\dagger)\ket{\mathbf 0}$, respectively. For a random non-interacting spin Hamiltonian $\hat H$ it is then sufficient to compute the Gram matrix $\mathbf G$ and the overlaps $\braket{\Psi_k|\hat \sigma^\alpha_l|\Psi}$, where $\alpha\in\{x,y,z\}$, $l\in\{1,2\}$ and $k\in\{1,2,3\}$ for the spin case and its analogue fermionic representation. We will only show the explicit results for the case $N=3$ and a specific representative. For instance, $\braket{\Psi_2|\hat \sigma^x_2|\Psi}=\eta_1\eta_1^*\eta_3\eta_3^* + \eta_1\eta_1^* + \eta_3\eta_3^* + 1$ and $\braket{\Phi_2|\hat \sigma^x_2|\Phi}=\xi_1\xi_1^*\xi_3\xi_3^* + \xi_1\xi_1^* + \xi_3\xi_3^* + 1$. One readily verifies that all Gram matrix elements and appearing operator overlaps also  coincide  under the identification $\eta_k = \xi_k$; We have verified all remaining elements but did not include them in this work for the sake of brevity. This can be naturally extended to larger spin system sizes $N>3$.  Consequently, the EOMs~\eqref{pr1} obtained from the two Ansätze~\eqref{pr4}--\eqref{pr5} are identical. This demonstrates that PV-FMFT exactly reproduces the SMFT dynamics for non-interacting spin-1/2 systems. We also note that the derivation of this Appendix can be extended to the ITE and can serve as an alternative proof to Appendix~\ref{product_state_construction_from_PVFGS}. 

\end{document}